\documentclass{aastex63}
\usepackage{amsmath}

\usepackage[frozencache,cachedir=.]{minted}
\newcommand{\fracbrac}[2]{\left(\frac{#1}{#2}\right)}
\newcommand{\dd}[2]{\frac{d#1}{d#2}}
\newcommand{\pd}[2]{\frac{\partial #1}{\partial #2}}

\newcommand{\ii}{\mathrm{i}}
\newcommand{\celmech}{\textsf{celmech}}
\newcommand{\rebound}{\textsf{REBOUND}}

\newcommand{\secref}[1]{Sec.\:\ref{#1}}
\newcommand{\paren}[1]{\left( #1\right)}
\newcommand{\bparen}[1]{\left[ #1\right]}

 \received{June 1,
2019} \revised{January 10, 2019} \accepted{\today} \submitjournal{AJ}
\shorttitle{the \textsf{celmech} code} \shortauthors{Hadden et al.}
\graphicspath{{./}{figures/}} \begin{document} 
\title{ \textsf{celmech}: A
\textsf{Python} Package for Celestial Mechanics }
\correspondingauthor{Sam~Hadden} \email{hadden@cita.utoronto.ca}
\author[0000-0002-1032-0783]{Sam~Hadden} \affiliation{Canadian Institute for
Theoretical Astrophysics, 60 St George St Toronto, ON M5S 3H8, Canada}
\author[0000-0002-9908-8705]{Daniel Tamayo} \affiliation{Department of Astrophysical Sciences,
Princeton University, Princeton, New Jersey 08544, USA} 
\begin{abstract} 
		We
		present \textsf{celmech}, an open-source \textsf{Python} package
		designed to facilitate a wide variety of celestial mechanics
		calculations.  The package allows users to formulate and integrate
		equations of motion incorporating user-specified terms from the classical
		disturbing function expansion of the interaction potential between
		pairs of planets.
		The code can be applied, for example, to isolate the contribution of particular resonances to a system's dynamical evolution and develop simple analytical models with the minimum number of terms required to capture a particular dynamical phenomenon.
		Equations and expressions can be easily manipulated by leveraging the
		extensive symbolic mathematics capabilities of the \textsf{sympy}
		\textsf{Python} package.  The \textsf{celmech} package is designed to
		interface seamlessly with the popular $N$-body code \textsf{REBOUND} to
		facilitate comparisons between calculation results and direct $N$-body
		integrations. The code is extensively documented and numerous example
		Jupyter notebooks illustrating its use are available online.
\end{abstract}

\section{Introduction}

\label{sec:intro} Celestial mechanics is the oldest
branch of physics and increasingly sophisticated analytical techniques for
predicting planetary motions have been developed over the course of the
subject's long history. While precise orbital solutions are now typically found
through direct numerical integrations of the equations of motion, analytic
methods remain central to solving open problems in dynamics.  We are often more
interested in developing theoretical insight into the qualitative elements
(e.g., mean motion resonances, secular evolution) responsible for dynamical
phenomena such as chaos, transit timing variations, extreme orbital
eccentricities etc.  This is especially true when studying exoplanetary
systems, where precise mass and orbital determinations are often unavailable,
and analytical intuition can guide the construction and interpretation of
suites of $N$-body integrations across a relevant subset of the vast parameter
space.  

The fact that planetary masses are much smaller than the central stellar mass,
and that orbital eccentricities and inclinations are typically small, often
makes it possible to model planetary dynamics  using only a small number of
slowly varying terms in the disturbing function expansion of the gravitational
interaction potential between planet pairs (a Fourier series in the angular
orbital elements and a power series in eccentricities and inclinations).
Nevertheless, this long-standing approach can be tedious when formulating even
relatively simple models as it entails identifying appropriate disturbing
function terms, finding expressions for their coefficients, and then evaluating
them by means of a variety of special functions. Examining any of the various
high-order expansion disturbing function expansions published during the 19th
century \citep[e.g.,][]{Peirce1849,LeVerrier1855,Boquet1889}, one quickly
appreciates that formulating any sophisticated, high-order analytic theories by
hand represents a monumental undertaking.  
 
Since the sixties, such detailed work has been performed using specialized
codes dedicated to computing and manipulating disturbing function expansions
\citep[see references in][]{Henrard1989,Brumberg1995}.  The state-of-the-art
TRIP code \citep[][]{Gastineau2011,TRIP}, a dedicated computer algebra system
for celestial mechanics, has proven especially valuable for studies the solar
system's long-term dynamics \citep[][]{Hoang2021,Hoang2022,Mogavero2022}.
However, such analyses have largely been limited to the specialized groups with
the expertise to develop these highly technical routines.\footnote{A limited
version of TRIP is available as a binary at
\href{https://www.imcce.fr/Equipes/ASD/trip/trip.php}{www.imcce.fr/Equipes/ASD/trip}.}

By contrast, the last decade has clearly demonstrated the impact open-source
scientific software can have on the field \citep{Tollerud2019}, both by opening
up specialized calculations to a wide base of users, and by providing
application programming interfaces (APIs) that enable the interconnection of
separate codes in a broad range of new and creative ways.  Particularly
successful cases in the exoplanet field include, e.g., \textsf{ttvfast}
\citep{Deck2014}, \textsf{EXOFAST} \citep[][]{Eastman2013}, \textsf{batman}
\citep{Kreidberg2015}, \textsf{radvel} \citep[][]{Fulton2018}, \textsf{REBOUND}
\citep[][]{ReinLiu2012}, \textsf{REBOUNDx} \citep[][]{reboundx},
\textsf{exostriker} \citep{Trifonov2019}, \textsf{NbodyGradient}
\citep[][]{Agol2021}, and \textsf{exoplanet} \citep{exoplanet:joss}.

\citet[][]{EllisMurray2000} provided an important early step towards making
computer-assisted disturbing function expansions more accessible to the wider
research community. They derived a method for calculating coefficients
associated with specific Fourier terms in the disturbing function. This offers
an efficient alternative to computing complete high-order disturbing function
expansions by means of dedicated computer algebra.  \textsf{Mathematica}
notebooks implementing their algorithm were made available as part of the
online material accompanying the popular textbook \citet[][]{MD1999ssd}.

In this paper, we present \textsf{celmech}, an open-source \textsf{Python}
package, which facilitates the manipulation of disturbing function expansions
to any order in the eccentricities and inclinations, and enables a broad range
of analytical investigations.  \textsf{celmech} is built upon on the disturbing
function expansion algorithm derived by \citet{EllisMurray2000}, though with
important modifications described in Section \ref{sec:disturbing_function} that
allow the equations of motion to be formulated in a powerful Hamiltonian
framework instead of in orbital elements.  We have designed the code's API to
easily interface with the \textsf{REBOUND} package, streamlining comparisons
between semi-analytic \celmech~models and direct $N$-body integrations.
Additionally, \celmech~includes functionality for performing canonical
transformations, which are useful for both extracting conserved quantities and
reducing the dimensionality of various problems
(Sec.\:\ref{sec:hamiltonian:canonical_transformations}).  We also plan to
incorporate additional modules in upcoming releases.

The plan of the paper is as follows:  we present a simple example application
in Section \ref{sec:example} with the goal of introducing readers to the code
and illustrating a typical workflow before proceeding through more in-depth
discussions of technical details in subsequent sections.  Section
\ref{sec:canonical_coordinates} reviews the coordinate system and Hamiltonian
framework that \textsf{celmech} uses to formulate the equations of motion.
Section \ref{sec:disturbing_function} describes \celmech's capabilities for
performing disturbing function expansions in terms of both classical Keplerian
orbital elements (Sec\:\ref{sec:disturbing_function:elements}), as well as
canonical variables (Sec\:\ref{sec:disturbing_function:canonical}).  Section
\ref{sec:hamiltonian} introduces \celmech's object-oriented approach to
modeling Hamiltonian systems in general (\secref{sec:hamiltonian:general}) as
well as those specific to planetary systems
(\secref{sec:hamiltonian:poincare}). It also describes \celmech's ability to
manipulate Hamiltonian models via the application canonical transformations
(\secref{sec:hamiltonian:canonical_transformations}), including Lie series
transformation series that arise in Hamiltonian perturbation theory
(\secref{sec:hamiltonian:canonical_transformations:lie-series}).  To aid the
reader, we provide a list the main mathematical symbols appearing in the paper
along with their definitions in Appendix \ref{app:list_of_symbols}.

\section{A Simple Example}
\label{sec:example}

We begin with a practical example to highlight some of the central machinery in
\celmech, demonstrate the code's API, and illustrate a possible workflow.  We
consider a system of three Earth-mass planets around a solar-mass star.  The
inner two are initialized at the 3:2 mean-motion resonance (MMR) with orbital
periods of $1$ and $1.5$ years respectively, while the third planet's orbital
period is 3.2 years and does not form any simple integer ratios with the inner
two.  The outermost orbit begins with zero eccentricity and inclination to the
reference plane.  The innermost orbit has both its eccentricity and inclination
(in radians) set to 0.02, while the middle orbit has these quantities both set
to 0.03.  The longitudes of ascending node, and the pericenters of the inner
two orbits lie along the positive $x$-axis, and both planets start at
conjunction at their respective apocenters where their orbits are spaced
furthest apart (Fig.\:\ref{fig:firstorder}).  The accompanying
\href{https://github.com/shadden/celmech/blob/master/jupyter_examples/AddingDisturbingFunctionTerms.ipynb}{Jupyter
notebook} shows how \celmech~conveniently interfaces with the \rebound~$N$-body
package, which provides extensive functionality for initializing orbital
configurations that can then be converted to a
\celmech~\textsf{PoincareHamiltonian} object. The \textsf{PoincareHamiltonian}
class, detailed below in Section\:\ref{sec:hamiltonian:poincare}, will be used
throughout the present example to build and manipulate a Hamiltonian model for
the system's dynamics.

\subsection{First-Order Approximation}

 The \textsf{PoincareHamiltonian} class has several  functions for adding
 specific terms from the disturbing function between pairs of planets in a
 system that can be used build up a dynamical model.  We expect that the 3:2
 MMR between the inner two planets will be important, so we call the following
 method of our \textsf{PoincareHamiltonian} object, which we named \textsf{Hp}:
	\begin{minted}{python}
	Hp.add_MMR_terms(p=3, q=1, indexIn=1, indexOut=2)
	\end{minted}
This adds the lowest order disturbing function terms in eccentricities and
inclinations that are associated with a 3:2 MMR between planets 1 and 2 (the
star has index 0) to the Hamiltonian represented by the
\textsf{PoincareHamiltonian} object \textsf{Hp}.  More generally, the arguments
\textsf{p} and \textsf{q} are used to specify terms associated with any the
$p:p-q$ resonance and the arguments \textsf{indexIn} and \textsf{indexOut}
specify the pair of planets for which terms should be added.

We can check the terms added to the disturbing function terms by inspecting the
\textsf{Hp.df} attribute, which prints them with their associated normalization
in units of energy (this disambiguation is useful in systems with more than one
pair of planets):
\begin{equation}
		-\frac{Gm_1m_2}{a_{2,0}}\tilde{C}_{(3,-2,0,-1,0,0)}^{(0,0,0,0),(0,0)}(\alpha_{1,2})e_2\cos(3\lambda_2-2\lambda_1-\varpi_2)
		-
		\frac{Gm_1m_2}{a_{2,0}}\tilde{C}_{(3,-2,-1,0,0,0)}^{(0,0,0,0),(0,0)}(\alpha_{1,2})e_1\cos(3\lambda_2-2\lambda_1-\varpi_1)
		\label{firstorderdf} 
\end{equation}
At first order in eccentricities and inclinations there are only two disturbing
function terms associated with the 3:2 MMR with the resonant angles
$3\lambda_2-2\lambda_1-\varpi_1$ and $3\lambda_2-2\lambda_1-\varpi_2$. These
terms would typically be found by hand, e.g., in the Appendix of
\cite{MD1999ssd}. In the notation\footnote{We modify the notation of
\citet[][]{MD1999ssd} slightly by subscripting the inner and outer planet's
variables by their indices, and labeling their $f_i$ as $f^{(j)}_{i}(\alpha)$
to make explicit the coefficients' dependence on the semi-major axis ratio
$\alpha_{1,2}=a_1/a_2$ and the coefficient, $j$, multiplying the outer planet's
mean longitude, $\lambda_2$, in the cosine argument.} of \cite{MD1999ssd}, the
expression would read
\begin{eqnarray}
    &\mathrm{MD:}& -\frac{Gm_1m_2}{a_{2,0}}\Bigg[f^{(3)}_{27}(\alpha_{1,2})e_1\cos(3\lambda_2-2\lambda_1-\varpi_1)
    +
    f^{(3)}_{31}(\alpha_{1,2})e_2\cos(3\lambda_2-2\lambda_1-\varpi_2)\Bigg]~.
\end{eqnarray}
Our $\tilde{C}$ coefficients correspond to the $f$ coefficients of
\cite{MD1999ssd}, where we choose to include some additional granularity in our
$\tilde{C}$ indices for specifying higher order terms.  The index convention is
explained in the following subsection (Sec.\:\ref{sec:secondorder}). 

\begin{figure*}
    \centering \resizebox{\columnwidth}{!}{\includegraphics{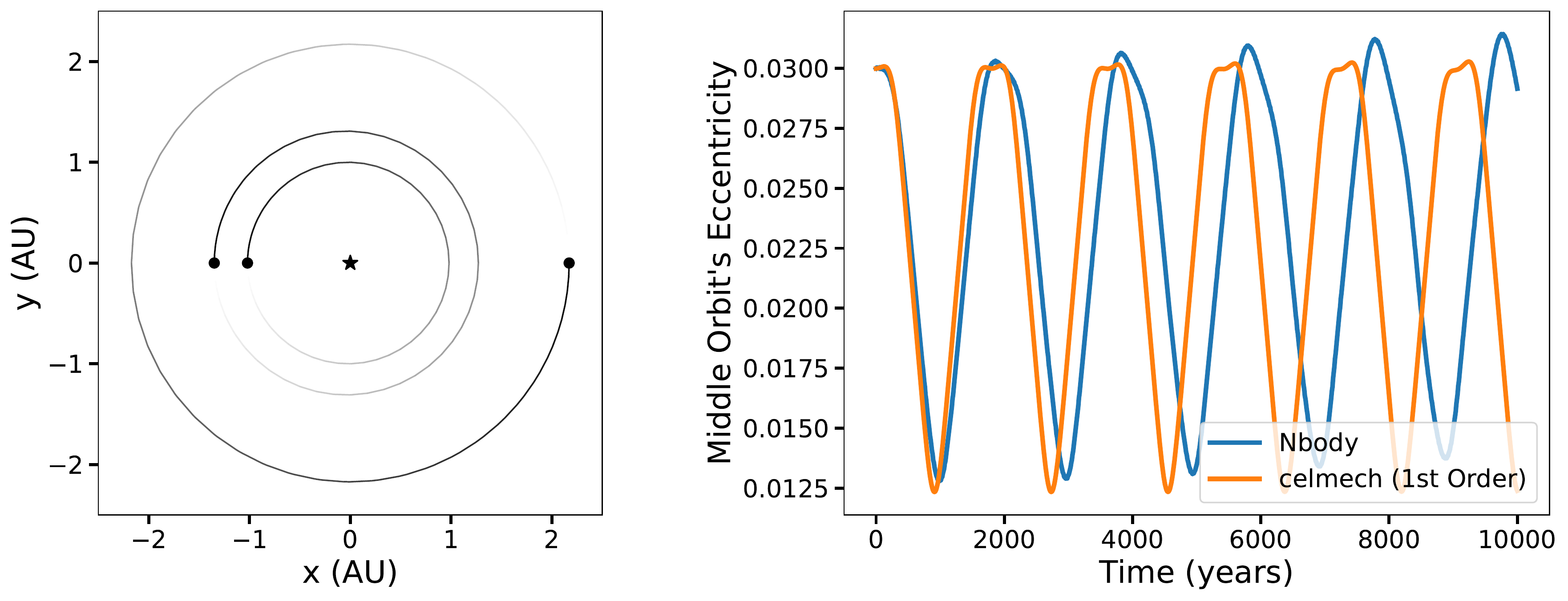}}
	\caption{Left panel: Initial orbital configuration in the reference $x-y$
		plane (inclinations to this plane are small $\lesssim 2^\circ$). Right
		panel: Comparison of the evolution of the middle planet's orbital
		eccentricity between a \celmech~model including the leading-order 3:2
		MMR terms in the disturbing function between the inner two planets
		(orange, Equation \eqref{firstorderdf}) to an $N$-body integration
		(blue).  \label{fig:firstorder}}
\end{figure*}

The \textsf{PoincareHamiltonian} object also allows us to integrate the
equations of motion using only the terms one has added to the disturbing
function.  We show the evolution for the middle planet's orbital eccentricity
in Fig.\:\ref{fig:firstorder}, comparing an $N$-body integration using
\rebound~to our simple \celmech~model incorporating only the leading order
resonant terms between the inner two planets.

We see that the oscillation amplitude and frequency is reproduced to within
$\approx 10\%$, though this causes the two models to quickly lose phase
coherence.  For many applications this level of agreement is sufficient, but we
can improve the agreement by incorporating more terms at higher order in the
eccentricities and inclinations.

\subsection{Going to Second Order} \label{sec:secondorder}

Even going to just the  next (second) order in the eccentricities and
inclinations introduces a large number of additional terms. The disturbing
function terms required to account for the 3:2 MMR interactions between the
inner planet pair and the secular interactions between all planet pairs at
second order in eccentricities and inclinations are given by
\begin{equation}
\begin{split}
&-\frac{Gm_1m_2}{a_{2,0}}\tilde{C}_{(3,-2,0,-1,0,0)}^{(0,0,0,0)}(\alpha_{1,2})e_2\cos(3\lambda_2-2\lambda_1-\varpi_2) 
-\frac{Gm_1m_2}{a_{2,0}}\tilde{C}_{(3,-2,-1,0,0,0)}^{(0,0,0,0)}(\alpha_{1,2})e_1\cos(3\lambda_2-2\lambda_1-\varpi_1) \\
&-\frac{Gm_1m_2}{a_{2,0}}\tilde{C}_{(6,-4,0,-2,0,0)}^{(0,0,0,0)}(\alpha_{1,2})e_2^2\cos(6\lambda_2-4\lambda_1-2\varpi_2)
-\frac{Gm_1m_2}{a_{2,0}}\tilde{C}_{(6,-4,-1,-1,0,0)}^{(0,0,0,0)}(\alpha_{1,2})e_1e_2\cos(6\lambda_2-4\lambda_1-\varpi_1-\varpi_2) \\
&-\frac{Gm_1m_2}{a_{2,0}}\tilde{C}_{(6,-4,-2,0,0,0)}^{(0,0,0,0)}(\alpha_{1,2})e_1^2\cos(6\lambda_2-4\lambda_1-2\varpi_1)
-\frac{Gm_1m_2}{a_{2,0}}\tilde{C}_{(6,-4,0,0,0,-2)}^{(0,0,0,0)}(\alpha_{1,2})i_2^2\cos(6\lambda_2-4\lambda_1-2\Omega_2) \\
&-\frac{Gm_1m_2}{a_{2,0}}\tilde{C}_{(6,-4,0,0,-1,-1)}^{(0,0,0,0)}(\alpha_{1,2})i_1i_2\cos(6\lambda_2-4\lambda_1-\Omega_1-\Omega_2)
-\frac{Gm_1m_2}{a_{2,0}}\tilde{C}_{(6,-4,0,0,-2,0)}^{(0,0,0,0)}(\alpha_{1,2})i_1^2\cos(6\lambda_2-4\lambda_1-2\Omega_1) \\
&-\frac{Gm_1m_2}{a_{2,0}}\tilde{C}_{(0,0,0,0,0,0)}^{(0,0,0,1)}(\alpha_{1,2})e_2^2
-\frac{Gm_1m_2}{a_{2,0}}\tilde{C}_{(0,0,0,0,0,0)}^{(0,0,1,0)}(\alpha_{1,2})e_1^2
-\frac{Gm_1m_2}{a_{2,0}}\tilde{C}_{(0,0,0,0,0,0)}^{(0,1,0,0)}(\alpha_{1,2})i_2^2
-\frac{Gm_1m_2}{a_{2,0}}\tilde{C}_{(0,0,0,0,0,0)}^{(1,0,0,0)}(\alpha_{1,2})i_1^2 \\
&-\frac{Gm_1m_2}{a_{2,0}}\tilde{C}_{(0,0,-1,1,0,0)}^{(0,0,0,0)}(\alpha_{1,2})e_1e_2\cos(-\varpi_1+\varpi_2)
-\frac{Gm_1m_2}{a_{2,0}}\tilde{C}_{(0,0,0,0,-1,1)}^{(0,0,0,0)}(\alpha_{1,2})i_1i_2\cos(-\Omega_1+\Omega_2) \\
&-\frac{Gm_1m_3}{a_{3,0}}\tilde{C}_{(0,0,0,0,0,0)}^{(0,0,0,1)}(\alpha_{1,3})e_3^2
-\frac{Gm_1m_3}{a_{3,0}}\tilde{C}_{(0,0,0,0,0,0)}^{(0,0,1,0)}(\alpha_{1,3})e_1^2
-\frac{Gm_1m_3}{a_{3,0}}\tilde{C}_{(0,0,0,0,0,0)}^{(0,1,0,0)}(\alpha_{1,3})i_3^2
-\frac{Gm_1m_3}{a_{3,0}}\tilde{C}_{(0,0,0,0,0,0)}^{(1,0,0,0)}(\alpha_{1,3})i_1^2 \\
&-\frac{Gm_1m_3}{a_{3,0}}\tilde{C}_{(0,0,-1,1,0,0)}^{(0,0,0,0)}(\alpha_{1,3})e_1e_3\cos(-\varpi_1+\varpi_3)
-\frac{Gm_1m_3}{a_{3,0}}\tilde{C}_{(0,0,0,0,-1,1)}^{(0,0,0,0)}(\alpha_{1,3})i_1i_3\cos(-\Omega_1+\Omega_3) \\
&-\frac{Gm_2m_3}{a_{3,0}}\tilde{C}_{(0,0,0,0,0,0)}^{(0,0,0,1)}(\alpha_{2,3})e_3^2
-\frac{Gm_2m_3}{a_{3,0}}\tilde{C}_{(0,0,0,0,0,0)}^{(0,0,1,0)}(\alpha_{2,3})e_2^2
-\frac{Gm_2m_3}{a_{3,0}}\tilde{C}_{(0,0,0,0,0,0)}^{(0,1,0,0)}(\alpha_{2,3})i_3^2
-\frac{Gm_2m_3}{a_{3,0}}\tilde{C}_{(0,0,0,0,0,0)}^{(1,0,0,0)}(\alpha_{2,3})i_2^2 \\
&-\frac{Gm_2m_3}{a_{3,0}}\tilde{C}_{(0,0,-1,1,0,0)}^{(0,0,0,0)}(\alpha_{2,3})e_2e_3\cos(-\varpi_2+\varpi_3)
-\frac{Gm_2m_3}{a_{3,0}}\tilde{C}_{(0,0,0,0,-1,1)}^{(0,0,0,0)}(\alpha_{2,3})i_2i_3\cos(-\Omega_2+\Omega_3)~.
\end{split} 
\label{eq:secorder}
\end{equation}
We have gone from just two terms in Equation \eqref{firstorderdf} to 26 terms
in Equation \eqref{eq:secorder}.  While the coefficients of such an expansion
would be extremely tedious to look up by hand, these terms are easy to
incorporate with \celmech. 

As can be verified in Equation \eqref{eq:secorder}, the $\tilde{C}$ subscript
indices specify the coefficients of the cosine arguments
\begin{equation}
(k_1, k_2, k_3, k_4, k_5, k_6) \longleftrightarrow \cos (k_1 \lambda_{out} + k_2 \lambda_{in} + k_3 \varpi_{in} + k_4 \varpi_{out} + k_5 \Omega_{in} + k_6 \Omega_{out}), \label{eq:kvec}
\end{equation}
where the ``in" and ``out" subscripts refer to the inner and outer body,
respectively, and the ordering of the $k$ coefficients has been chosen to match
the ordering conventionally used in the cosine arguments
\citep[e.g.,][]{MD1999ssd}.  The amplitude of each cosine term is a power
series in the eccentricities and inclinations.  We choose to separate the terms
in these expansions, and specify them by the $\tilde{C}$ superscript indices,
as detailed in Sec.\:\ref{sec:disturbing_function:canonical}.  Most terms in
Equation \eqref{eq:secorder} are the leading order contribution with
superscripts $(0,0,0,0)$, though we see examples of higher order contributions
in the secular terms without a cosine argument (all $k_i=0$ in Equation
\ref{eq:kvec}).

The most important second-order terms for our setup are the second-order 3:2
MMR terms, i.e., the 6:4 cosine terms above, involving the combination
$6\lambda_2 - 4\lambda_1$.  In \celmech, we would incorporate MMR terms up to a
specified maximum order by adding the \textsf{max\_order} keyword to our call
above:
	\begin{minted}[linenos]{python}
	Hp.add_MMR_terms(p=3, q=1, max_order=2, indexIn=1, indexOut=2)
	\end{minted}

The secular terms (terms above that do not depend on the mean longitudes
$\lambda$, including those with no cosine factors) also make a small, but
noticeable impact.  For each pair of planet indices \textsf{(idx1, idx2)}, the
secular terms can be added by
	\begin{minted}[linenos]{python}
	Hp.add_secular_terms(max_order=2, indexIn=idx1, indexOut=idx2) 
	\end{minted}

\begin{figure}
    \centering \resizebox{\columnwidth}{!}{\includegraphics{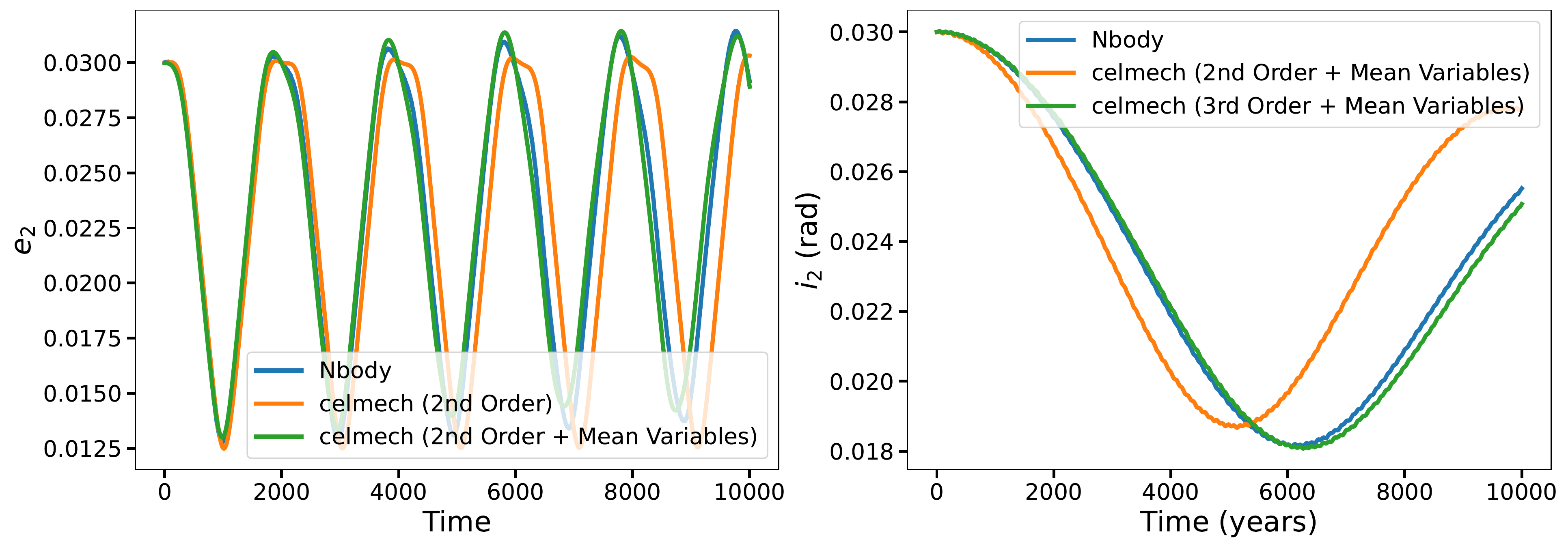}}
	\caption{Left panel: Comparison of the evolution of the middle orbit's
		eccentricity in our example between $N$-body (blue) and increasingly
		sophisticated \celmech~models. The orange curve incorporates the most
		important terms to second order in the eccentricities and inclinations,
		while the green curve additionally incorporates a correction from
		osculating to mean variables (see text). Right panel: Comparison of the
		evolution of the evolution of the middle orbit's inclination between
		$N$-body (blue) and a second-order (orange) and third-order (green)
		\celmech~ model (both corrected to mean variables).
		\label{fig:higherorder}}
\end{figure}

Comparing Fig.\:\ref{fig:firstorder} with the left panel of
Fig.\:\ref{fig:higherorder}, we see that this second-order model (orange curve)
does improve the agreement with $N$-body.

\subsection{Mean Variables}

While it is tempting to simply continue adding higher order terms, in this case
it would not help.  There is a separate issue causing a discrepancy with the
$N$-body integrations, which can be fixed (at negligible computational cost) by
converting from ``osculating" to ``mean" variables.  Any time one takes only a
subset of terms in the disturbing function, one is considering a (slightly)
different problem from the full one.  This distinction is particularly
important, given our rather special initial conditions. 

Each time an inner planet overtakes its outer neighbor at conjunction, it gains
orbital energy on approach as it is being pulled forward by its outer neighbor,
only to lose that energy approximately symmetrically on the way out.  These
encounters cause short-period ``spikes" in the orbits' semimajor axes, and the
fact that we started our planets at conjunction, means that we are starting at
the peak of one of those ``spikes".  In other words, our initial semimajor axes
are in a sense, spurious, and not reflective of the mean values they have over
the vast majority of the orbit.  This in turn impacts the period ratio between
the inner two planets and their resonant dynamics.

Rather than incorporating even more terms in our disturbing function, the most
elegant way of handling these (and other) ignored terms is through a
near-identity canonical transformation from ``osculating" to ``mean" variables
using Lie series (\citealt{Morbidelli2002},
Sec.\:\ref{sec:hamiltonian:canonical_transformations:lie-series}).  One can do
this in \celmech~by creating a \textsf{FirstOrderGeneratingFunction} object,
and specifying the terms one wants to average out.  The terms in the disturbing
function causing the semimajor axis spikes discussed above are of the form
$\cos\left[k(\lambda_2-\lambda_1)\right]$, and can be analytically summed over
(\citealt{Malhotra1993},
Sec.\:\ref{sec:hamiltonian:canonical_transformations:lie-series}).  These terms
are independent of the eccentricities and inclinations, and thus of
zeroth-order, and can be added in \celmech~starting from a set of
\textsf{poincare\_variables} specifying the (osculating) initial conditions via 
	\begin{minted}[linenos]{python}
	chi = FirstOrderGeneratingFunction(poincare_variables)
	chi.add_zeroth_order_term()
	\end{minted}
The effects of unmodeled, nearby MMRs are much smaller for this problem than
the zeroth-order terms above, but still make a noticeable difference, and are
added by calls of the form 
	\begin{minted}[linenos]{python}
	chi.add_MMR_terms(p=2,q=1,l_max=1, indexIn=1, indexOut=2)
	chi.add_MMR_terms(p=4,q=1,l_max=1, indexIn=1, indexOut=2)
	\end{minted}
Once we have incorporated all the terms for which we want to correct, we
transform to mean variables with \textsf{chi.osculating\_to\_mean()}, and the
resulting integration (green curve in left panel of
Fig.\:\ref{fig:higherorder}), yields excellent agreement with
$N$-body.

\subsection{Going to Higher Order}

In the right panel of Fig.\:\ref{fig:higherorder}, we can see that, even at
second-order (orange), there is a significant discrepancy with $N$-body (blue)
in the orbital inclination evolution of the middle planet.  In order to
approach the $N$-body evolution, we have to include terms up to third order in
the eccentricities and inclinations.  This is the order at which terms directly
coupling the inclinations to the quickly varying eccentricities first appear.
These are easily obtained by setting \textsf{max\_order=3} in the calls above,
though we spare the reader the details of the resulting 56 terms in the
disturbing function. The \textsf{add\_secular\_terms} method accepts a similar
\textsf{max\_order} keyword argument, but higher order secular terms do not
appear until 4th order in eccentricities and inclinations.

\subsection{Limits}

The degree of agreement between $N$-body and \celmech~models will depend on
planet masses, with smaller masses generally giving better agreement.  In
particular, the conversion from ``osculating" to ``mean" variables is not
perfect.  The Lie series transformation
(Sec.\:\ref{sec:hamiltonian:canonical_transformations:lie-series}) corrects
unmodeled terms (e.g., conjunctions, nearby MMRs) to first order in the
planetary masses.  This generates second-order (and higher) mass terms that
should be added to the equations of motion governing the evolution of the
``mean" variables.  For our low-mass planets these corrections are negligible,
but at higher mass-ratios with the central star, these terms start to be
significant \citep{SansotteraLibert2019}.  Adding higher order terms in the
eccentricities and inclinations will help until one reaches the level at which
these second-order mass terms become significant.

\subsection{Applications}

Consideration of only the leading order terms is attractive for both analytical
understanding and computational cost.  As one incorporates more terms of
progressively higher order, \celmech~can become slower than direct $N$-body
integrations.  Nevertheless, there are several applications for high-order
models.  First, in many instances it may be possible to deploy integration
algorithms that are more efficient than \celmech's  default general-purpose
integrator and any such algorithms can potentially make use lower-level
derivative and Jacobian information generated by the
\textsf{PoincareHamiltonian} instance (i.e., \textsf{Hp}).  This derivative and
Jacobian information can also be valuable for locating equilibrium
configurations such as periodic orbits via root-finding methods.  High-order
expansion can also be the starting point for developing accurate integrable
approximations of the dynamics through canonical transformations
\citep{Hadden2019}.  Finally, incremental comparisons of successively higher
order expansions with $N$-body integrations can also provide a quick guide to
the degree of complexity required to model a particular dynamical phenomenon of
interest.

\section{Coordinate System}
\label{sec:canonical_coordinates}

This section details the coordinate system and dynamical variables the
\textsf{celmech} code uses for formulating Hamiltonian equations of motion. We
begin by reviewing the Hamiltonian formulation of the planetary $N$-body
problem, considering $N-1$ planets of mass $m_i$ with $i=1,...,N-1$ orbiting a
central star of mass $M_*$.  Let $\pmb{u}_*$ and $\pmb{u}_i$ be the Cartesian
position vectors of the star and the planets, respectively, in the inertial
barycentric frame.  We start by formulating the Hamiltonian in term of
canonical heliocentric coordinates, $(\pmb{r}_i,\tilde{\pmb{r}}_i)$, where
$\pmb{r}_i = \pmb{u}_i -\pmb{u}_*$ is the  $i$th planet's heliocentric position
vector and $\tilde{\pmb{r}}_i = m_i\dot{\pmb{u}}_i$ is its canonically
conjugate momentum \citep{Laskar1995}. In  these coordinates, the Hamiltonian
reads \begin{eqnarray} H &=& \sum_{i=1}^{N-1}H_{\mathrm{Kep},i} +
\sum_{i=1}^{N-1} \sum_{j=1}^{i} H_{\mathrm{int.}}^{(i,j)} \end{eqnarray} where
\begin{eqnarray} H_{\mathrm{Kep},i} &=&
		\frac{1}{2}\frac{|\tilde{\pmb{r}}_i|^2}{\mu_i}   -
		\frac{GM_i\mu_i}{|\pmb{r}_i|}
		\label{eq:hamiltonian:Hkep_canonical_heliocentric} \\
		H_{\mathrm{int.}}^{(i,j)} &=& -\frac{Gm_im_j}{|\pmb{r}_i - \pmb{r}_j|}
		+ \frac{\tilde{\pmb{r}}_i \cdot \tilde{\pmb{r}}_j }{M_*}~
		\label{eq:hamiltonian:Hint_canonical_heliocentric}, 
\end{eqnarray}
${\mu}_i = \frac{m_iM_*}{M_* + m_i}$, and ${M}_i = {M_* +
m_i}$.\footnote{
		Strictly speaking, Hamiltonian
		\eqref{eq:hamiltonian:Hkep_canonical_heliocentric} should also include the term
		$- \frac{\tilde{\pmb{r}}_0}{M_*} \cdot \sum_{i \ne 0} \tilde{\pmb{r}}_i$ where
		$\tilde{\pmb{r}}_0$ is the total momentum of the system and is conjugate to the
		canonical coordinate $\pmb{r}_0=\pmb{u}_*$ \citep[see, e.g., Equation 27 of
		][]{HernandezDehnen2017}. This term is necessary to derive the correct
		canonical equation of motion for $\pmb{r}_0$ but is not needed to get the
		correct canonical equations for $\pmb{r}_i$ with $i>0$ because
		$\tilde{\pmb{r}}_0\equiv 0$. The time-evolution of $\pmb{r}_0$ can be
		calculated without explicitly integrating its equation of motion by noting that
		$\pmb{r}_0 =-\sum_{i=1}^{N-1} m_i \pmb{r}_i / (M_*  + \sum_{i=1}^{N-1} m_i)$.
}
 
We will perform a canonical transformation from the canonical heliocentric
coordinates  $(\pmb{r}_i,\tilde{\pmb{r}}_i)$ to the canonical modified Delaunay
variables that constitute action-angle coordinates of the integrable two-body
Hamilontians, $H_{\mathrm{Kep},i}$.  In order to effect this transformation, we
first define a set of `canonical heliocentric' Keplerian orbital elements,
$(a_i,e_i,I_i,\lambda_i,\varpi_i,\Omega_i)$, which represent, respectively, the
planetary orbit's semi-major axis, eccentricity, inclination, mean longitude,
longitude of periapse, and longitude of ascending node.  Standard orbital
elements are computed from bodies' three-dimensional Cartesian positions and
velocities as well as a ``gravitational parameter"  \citep[e.g.,][]{MD1999ssd}.
For each planet, canonical heliocentric elements are computed from the
Cartesian position vector $\pmb{r}_i$, the ``velocity" vector
$\tilde{\pmb{r}}_i/\mu_i$, and the gravitational parameter $GM_i$. While
$\tilde{\pmb{r}}_i/\mu_i$  does not correspond to any physical velocity vector
in the system, this definition of canonical heliocentric elements ensures that
the modified Delaunay variables, defined below, form a canonical set of
action-angle coordinates for the two-body Hamiltonian, $H_{\mathrm{Kep},i}$.
The \celmech~module {\textsf{nbody\_simulation\_utilities}} provides the
function \textsf{reb\_calculate\_orbits} to compute canonical heliocentric
orbital elements from a \rebound~simulation and the function
\textsf{reb\_add\_from\_elements} to add particles to a \rebound~simulation by
specifying these elements
(\href{https://github.com/shadden/celmech/blob/master/jupyter_examples/CoordinatesConvertingToFromNbody.ipynb}{Jupyter
Notebook example}).

The canonical modified Delaunay coordinates\footnote{Different literature
sources use different names for these canonical variables.  \citet{MD1999ssd}
refer to these variables as ``Poincaré" variables whereas
\citet{Morbidelli2002} refers to these variables as modified Delaunay variables
and reserves the name ``Poincaré" variables for the set of canonical variables
introduced in Equation \eqref{eq:poincare_def}.  } are defined in terms of the
canonical heliocentric orbital elements. The action variables are
\begin{eqnarray} 
		\Lambda_i &=& \mu_i\sqrt{GM_ia_i}\nonumber\\ \Gamma_i  &=&
		\mu_i\sqrt{GM_ia_i}(1-\sqrt{1-e_i^2})\nonumber\\ Q_i &=&
2\mu_i\sqrt{M_ia_i}\sqrt{1-e_i^2}\sin^2(I_i/2)
\label{eq:hamiltonian:poincare:momenta} 
\end{eqnarray} 
and have units of angular momentum. The angle variables canonically conjugate
to the action variables $\Lambda_i,\Gamma_i,$ and $Q_i$ are $\lambda_i$,
$\gamma_i=-\varpi_i$, and $q_i=-\Omega_i$, respectively. In order to avoid
coordinate system singularities that leave the canonical variables $\gamma_i$
and $q_i$ undefined when $e_i=0$ and $I_i=0$, \celmech~formulates equations of
motion in terms canonical coordinate-momentum pairs \begin{eqnarray}
		\label{eq:poincare_def} (\eta_i,\kappa_i) &=& \sqrt{2\Gamma_i}\times
(\sin \gamma_i,\cos \gamma_i)\nonumber\\ (\rho_i,\sigma_i) &=&
\sqrt{2Q_i}\times (\sin q_i,\cos q_i) \end{eqnarray} rather than
$(\Gamma_i,\gamma_i)$ and $(Q_i,q_i)$.  In modified Delaunay variables, the
two-body Hamiltonian, Equation
\eqref{eq:hamiltonian:Hkep_canonical_heliocentric}, simply becomes
\begin{equation} \label{eq:hamiltonian:Hkep_def} H_{\mathrm{Kep},i} =
\frac{G^2M_i^2\mu_i^3}{2\Lambda_i^2}~.  \end{equation} By contrast,
$H_{\mathrm{int.}}^{(i,j)}$ does not admit a simple expression in terms of the
canonical variables. Rather, the benefit of expressing the Hamiltonian in terms
of classical action angle variables is that $H_{\mathrm{int.}}^{(i,j)}$ can be
decomposed as a Fourier series in the canonical angle variables, enabling the
application of methods of Hamiltonian perturbation theory.

\section{Expansion of the Disturbing Function}
\label{sec:disturbing_function}

The \textsf{celmech.disturbing\_function} module provides tools for
calculating coefficients appearing in the cosine series expansion of the
interaction Hamiltonian, $H_{\mathrm{int.}}^{(i,j)}$ both in terms of both
orbital elements as well as canonical variables. We begin by describing the
expansion of the disturbing function in terms of orbital elements.
    
\subsection{Expansion in Terms of Orbital Elements}
\label{sec:disturbing_function:elements} 

In order to write the cosine series expansion, we will take $a_j>a_i$, define
$\alpha = a_i/a_j$, $s_i=\sin(I_i/2)$, $\pmb{\theta}_{i,j} =
(\lambda_j,\lambda_i,-\gamma_i,-\gamma_j,-q_i,-q_j)$ and write
    \begin{equation}
        H^{(i,j)}_\mathrm{int}= 
        -\frac{Gm_im_j}{a_j}
        \sum_{\bf k}
        \sum_{\nu_1,\nu_2,\nu_3,\nu_4=0}^\infty 
          \tilde{C}_{\bf k}^{\pmb{
         \nu}}(\alpha_{i,j})s_i^{|k_5|+2\nu_1}s_j^{|k_6|+2\nu_2}e_i^{|k_3|+2\nu_3}e_j^{|k_4|+2\nu_4}\cos({\bf k}\cdot \pmb{\theta}_{i,j})
        \label{eq:disturbing_function:Hint_fourier}
    \end{equation}
where  ${\bf k}=(k_1,k_2,...k_6)$ and $\pmb{\nu} = (\nu_1,...,\nu_4)$ are
integer multi-indices and $\alpha_{i,j}=a_i/a_j$.  Rotation and reflection
symmetries of the planets' gravitational interactions dictate that
$\tilde{C}^{\pmb{\nu}}_{\bf k}(\alpha)\ne 0 $ only if $\sum_{l=1}^{6}k_l = 0$
and $k_5+k_6=2n$ where $n$ is an integer \citep[e.g.,][]{MD1999ssd}. For small
eccentricities and inclinations, the coefficient of the cosine term $\cos({\bf
k}\cdot \pmb{\theta}_{i,j})$ is of order $\sim
s_i^{|k_5|}s_j^{|k_6|}e_i^{|k_3|}e_j^{|k_4|}$ and higher order corrections are
of degree 2 and greater in the planets' eccentricities and inclinations.
    
To compute the coefficients $\tilde{C}_{\bf k}^{\pmb{\nu}}(\alpha)$ appearing
in Equation \eqref{eq:disturbing_function:Hint_fourier}, we first represent the
interaction Hamiltonian as the sum of two parts, $H^{(i,j)}_\mathrm{int} =
-\frac{Gm_im_j}{a_j}({\cal R}^{(i,j)}_\mathrm{dir.} + {\cal
R}^{(i,j)}_\mathrm{ind.})$ where we define the direct and indirect parts of the
disturbing function as\footnote{
	Note that our definition of the indirect disturbing function,
	${\cal R}_\mathrm{ind.}$, differs from \citet{MD1999ssd} because we
	formulate our equations of motion in terms of canonical heliocentric
	variables.  \citet{MD1999ssd} formulate equations in terms of heliocentric
	position and velocity vectors that do not constitute canonically conjugate
	variable pairs. Consequently, their formulation of the equations of motion
	requires different indirect terms to be computed depending on whether a
	planet is subject to an interior or exterior perturber.
   }
    \begin{eqnarray}
        {\cal R}^{(i,j)}_\mathrm{dir.} &=& \frac{a_j}{|\pmb{r}_j - \pmb{r}_i|} \nonumber\\
        \mathrm{and}~ {\cal R}^{(i,j)}_\mathrm{ind.} &=& -\frac{a_j}{GM_*m_im_j}\tilde{\pmb{r}}_i \cdot \tilde{\pmb{r}}_j~
        \label{eq:disturbing_function:df_dir_ind}
    \end{eqnarray}
and we denote the contributions of ${\cal R}^{(i,j)}_\mathrm{dir.}$ and
${\cal R}^{(i,j)}_\mathrm{ind.}$ to the value of the coefficient
$\tilde{C}_{\bf k}^{\pmb{\nu}}(\alpha)$ as $[\tilde{C}_{\bf
k}^{\pmb{\nu}}(\alpha)]_\mathrm{ind.}$ and $[\tilde{C}_{\bf
k}^{\pmb{\nu}}(\alpha)]_\mathrm{ind.}$, respectively.  Given values of
$\bf{k}$ and $\pmb{\nu}$, the \celmech~code implements the algorithm
described in \citet[][]{EllisMurray2000} to compute $[\tilde{C}_{\bf
k}^{\pmb{\nu}}(\alpha)]_\mathrm{dir.}$ in terms of Laplace coefficients and
their derivatives while the calculation of $[\tilde{C}_{\bf
k}^{\pmb{\nu}}(\alpha)]_\mathrm{ind.}$ is detailed in Appendix
\ref{app:indirect}.  The \textsf{celmech.disturbing\_function} module's
\textsf{df\_coefficient\_Ctilde} function can be used to compute the
coefficients $\tilde{C}_{\bf k}^{\pmb{\nu}}(\alpha)$. The function takes
two integer tuples, $\textsf{k} = (k_1,k_2,...,k_6)$ and
$\textsf{nu}=(\nu_1,...,\nu_4)$, as arguments and returns a representation
of the coefficient as a \textsf{python} dictionary with key-value pairs in
the form $\{(p_1,(j_1,s_1,n_1)):A_1,...,(p_N,(j_N,s_N,n_N)):A_N,
(\mathtt{'indirect'},p_\mathrm{ind}):A_\mathrm{ind}\}$ with 
	\begin{equation}
		 \tilde{C}_{\bf k}^{\pmb{\nu}}(\alpha) = A_\mathrm{ind}\alpha^{-p_\mathrm{ind}/2} +  \sum_{l=1}^N   
		 A_l \alpha^{p_l}\frac{d^{n_l}}{d\alpha^{n_l}}b_{s_l}^{(j_l)}(\alpha)
		\label{eq:disturbing_function:laplace_coeff_sum}
	\end{equation}
where 
    \begin{eqnarray*}
    	 b_{s}^{(j)}(\alpha) = \frac{1}{\pi}\int_{-\pi}^\pi\frac{\cos(j\theta)}{(1+\alpha^2-2\alpha\cos\theta)^{s}}d\theta~,
    \end{eqnarray*}
$p_l, n_l,$ and $j_l$ are integers, and $s_l$ are half-integers.

The function \textsf{celmech.disturbing\_function} module's
\textsf{evaluate\_df\_coefficient\_dict} can be used to compute a numerical
value of a coefficient, $\tilde{C}_{\bf k}^{\pmb{\nu}}$, for a particular
choice of $\alpha$.  The function \textsf{deriv\_df\_coefficient} computes
derivatives of disturbing function coefficients with respect to $\alpha$.
This function takes a dictionary of the form returned by
\textsf{df\_coefficient\_Ctilde} as an argument and returns a new
dictionary of the same form representing the derivative of the coefficient
with respect to $\alpha$. Some basic calculations of disturbing function
coefficients are demonstrated in Jupyter example notebook
\href{https://github.com/shadden/celmech/blob/master/jupyter_examples/DisturbingFunctionCoefficients.ipynb}{DisturbingFunctionCoefficients.ipynb}.

\subsection{Expansion in Terms of Canonical Variables}
\label{sec:disturbing_function:canonical}

While our  $\tilde{C}_{\bf k}^{\pmb{\nu}}$ coefficients for the direct part of
the disturbing function match those of \citet[][]{EllisMurray2000} derived in
terms of orbital elements, formulating Hamiltonian equations of motion requires
a disturbing function expansion expressed in terms of canonical variables. To
derive an expansion in terms of canonical variables we first introduce some
auxillary variables: we introduce reference semi-major axes $a_{i,0}$ about
which expansion of coefficients will be centered along with $\alpha_{ij,0} =
a_{i,0}/a_{j,0}$, $\Lambda_{i,0} = \Lambda_{i}\sqrt{a_{i,0}/a_i}$ and $\delta_i
= (\Lambda_i - \Lambda_{i,0})/\Lambda_{i,0}$. Next, we define  $X_i =
\sqrt{\frac{1}{\Lambda_{i,0}}}(\kappa_i - \ii \eta_i)$ and $Y_i =
\frac{1}{2}\sqrt{\frac{1}{\Lambda_{i,0}}}(\sigma_i-\ii\rho_i)$, where we use a
Roman `$\ii$' to denote the imaginary unit, i.e., $\ii \equiv \sqrt{-1}$.  The
complex variables $z_i = e_ie^{\ii \varpi_i}$ and $\zeta_i = s_ie^{\ii
\Omega_i}$ can then be expressed explicitly in terms of the variables
$X_i,Y_i,$ and $\delta_i$ according to
    \begin{eqnarray}
        z_i &=& \frac{1}{\sqrt{1+\delta_i}}X_{i}\sqrt{1-\frac{1}{4(1+\delta_i)}X_i\bar{X}_i} 
        \label{eq:z_in_XY}
        \\
        \zeta_i&=&
        \frac{1}{\sqrt{1+\delta_i}}
        \frac{Y_i}
        {\sqrt{
    1-
    \frac{1}{2(1+\delta_i)}
    X_i\bar{X}_i}
    }~
    \label{eq:zeta_in_XY}
\end{eqnarray}
where overlines denote complex conjugates. To express the amplitude of a single
disturbing function cosine term, $\cos({\bf k}\cdot \pmb{\theta}_{i,j})$, in
terms of the canonical variables introduced in Section
\ref{sec:canonical_coordinates}, we seek an expression for coefficients $C_{\bf
k}^{\pmb{\nu},\pmb{l}}$  such that, when $k_3,k_4,k_5,$ and $k_6>0$, the
following relationship hold:
\begin{equation}
        \fracbrac{1}{a_j}
        z_i^{k_3}z_j^{k_4}\zeta_i^{k_5}\zeta_j^{k_6}
        \sum_{\pmb{\nu}=0}^\infty 
          \tilde{C}_{\bf k}^{\pmb{\nu}}(\alpha_{ij})
          s_i^{2\nu_1}s_j^{2\nu_2}e_i^{2\nu_3}e_j^{2\nu_4}
          =
        \fracbrac{1}{a_{j,0}}
        X_i^{k_3}X_j^{k_4}Y_i^{k_5}Y_j^{k_6}
        \sum_{\pmb{l}=0}^\infty
        \sum_{\pmb{\nu}=0}^\infty 
        {C}_{\bf k}^{\pmb{\nu},\pmb{l}}(\alpha_{ij,0})
        |Y_i|^{2\nu_1}|Y_j|^{2\nu_2}|X_i|^{2\nu_3}|X_j|^{2\nu_4}\delta_{i}^{l_1}\delta_{j}^{l_2}
        \label{eq:hamiltonian:poincare:complex_expansion}~.
\end{equation}
When $k_3<0$, $z_i^{k_3}$ and $X_i^{k_3}$ in Equation
\eqref{eq:hamiltonian:poincare:complex_expansion} are replaced with
$\bar{z}_i^{-k_3}$ and $\bar{X}_i^{-k_3}$, respectively, and analogous
substitutions for variables $(z_j,\zeta_i,\zeta_j)$ and  $(X_j,Y_i,Y_j)$ are
made when any of $k_4,k_5,$ or $k_6<0$.  In Appendix
\ref{app:canonical_expansion} we work out explicit expressions for the
coefficients ${C}_{\bf k}^{\pmb{\nu},\pmb{l}}$ in terms of terms of the
$\tilde{C}_{\bf k}^{\pmb{\nu}}$ coefficients and their derivatives.  While
these expressions are, in general, complicated, note that ${C}_{\bf
k}^{(0,0,0,0),(0,0)} = \tilde{C}_{\bf k}^{(0,0,0,0)}$. The
\textsf{disturbing\_function} module's \textsf{df\_coefficient\_C} function can
be used to obtain disturbing function coefficients ${C}_{\bf
k}^{\pmb{\nu},\pmb{l}}$.

\subsection{Generating disturbing function arguments of a given order}
\label{sec:disturbing_function:arguments}

In many applications, one is interested in gathering a series of particular
disturbing function terms, such as those associated with a particular
resonance, up to a maximum order, $N_\mathrm{max}$, in eccentricities and
inclinations. The \textsf{disturbing\_function} module provides the function
\textsf{df\_arguments\_dictionary} that allows the user to enumerate all cosine
arguments, ${\bf k}\cdot \pmb{\theta}_{ij}$, appearing in the disturbing
function expansion up to a given order $N_\mathrm{max}$ in inclinations and
eccentricities. In order to do so, the function enumerates tuples
$(k_3,k_4,k_5,k_6)$ that satisfy $|k_3| + |k_4| + |k_5| + |k_6| \le
N_\mathrm{max}$, along with the condition $k_5 + k_6 = 2n$ for integer $n$. The
latter condition ensures $C_{\bf k}^{\pmb{n},\pmb{l}}\ne0$ (see Section
\ref{sec:disturbing_function:elements}). These terms are further grouped by
their value of $\sum_{i=3}^6k_i$, which must be equal to $-(k_1  + k_2)$ to
ensure $C_{\bf k}^{\pmb{n},\pmb{l}}\ne0$.  The Jupyter notebook example
\href{https://github.com/shadden/celmech/blob/master/jupyter_examples/DisturbingFunctionArguments.ipynb}{\textsf{DisturbingFunctionArguments.ipynb}}
illustrates how the results returned by \textsf{df\_arguments\_dictionary} are
structured.

The \textsf{disturbing\_function} module also provides the convenience function
\textsf{list\_resonance\_terms} to create a list of all $\bf{k}$ and
$\pmb{\nu}$ combinations associated with a particular MMR that appear in the
expansion of the disturbing function up to a user-specified order. In
particular, given input $p$ and $q$ specifying the  $p$:$p-q$ along with
maximum order, \textsf{list\_resonance\_terms} returns a \textsf{python} list
of tuples in the form $((k_1,k_2,...,k_6),(\nu_1,...,\nu_4))$ that includes all
terms that satisfy
    \begin{eqnarray}
    &(q-p)k_1 + p k_2 = 0\nonumber\\
    &|k_3|+|k_4|+|k_5| + |k_6|+2(\nu_1+\nu_2+\nu_3+\nu_4)\le N_\mathrm{max}
    \end{eqnarray}
in addition to the usual rules $\sum_{i=1}^6k_i=0$ and $k_5+k_6=2n$ for integer
$n$.  Similarly, the function \textsf{list\_secular\_terms} can be used to
create a list of all $\bf{k}$ and $\pmb{\nu}$ combinations that appear as
secular terms with $k_1=k_2=0$ in the disturbing function expansion up to a
user-specified order.  Both the \textsf{list\_resonance\_terms} and
\textsf{list\_secular\_terms} functions are illustrated in a
\href{https://github.com/shadden/celmech/blob/master/jupyter_examples/DisturbingFunctionArguments.ipynb}{Jupyter
notebook example}.

\subsection{Oblateness and General Relativity}

The effects of primary oblateness and general relativistic precession can be
important for the dynamical evolution of planetary systems.  Here we give
expressions for the orbit-averaged potentials of these effects in terms of the
canonical variables used by \textsf{celmech}.  The gravitational potential due
to the quadrupole moment of an oblate central primary at a heliocentric
position $\mathbf{r}$ is given by
\begin{equation}
\phi_{J_2}(\mathbf{r}) =J_2\frac{GM_*R_*^2}{r^3}P_2(\hat{\mathbf{r}}\cdot\hat{\mathbf{z}})
\end{equation} 
where $r=|\mathbf{r}|$,  $\hat{\mathbf{r}} =\mathbf{r}/r$, and
$\hat{\mathbf{z}}$ is the direction perpendicular to the primary's equatorial
plane.  In terms of orbital elements, $\hat{\mathbf{r}}\cdot\hat{\mathbf{z}} =
\sin(I)\sin(f+\varpi-\Omega) = 2s\sqrt{1-s^2}\sin(f+\varpi-\Omega)$ where $f$
is the true anomaly and the inclination, $I$, is measured relative to the
primary's equatorial plane.  Denoting orbit-averaged values by $<\cdot>$, we
use $<r^{-3}> = a^{-3}\left(1-e^2\right)^{-3/2}$ and
$<r^{-3}\sin^2(f+\varpi-\Omega)> = \frac{1}{2}<r^{-3}>$ to find that the
orbit-averaged value of the quadrupole potential is given by $<\phi_{J_2}>=
-J_2\frac{GM_*}{a}\fracbrac{R_*}{a}^2{\left(1-e^2\right)^{-3/2}}\left(1/2+{3(s^4-
s^2)}\right)$.  The corresponding contribution to the Hamiltonian can be
written in terms of canonical variables as
\begin{eqnarray}
\label{eq:j2_potential}
       H_{J_2}(\Lambda,\eta,\kappa,\rho,\sigma)
       &=& -
       \frac{
            G^4M^4_*\mu^6J_2R_*^2
       }
       {
        \Lambda^3\left(\Lambda - \frac{1}{2}(\eta^2+\kappa^2)\right)^3
       }
       \left[
       \frac{1}{2} - 
       \frac{3}{4} \fracbrac{
                \rho^2+\sigma^2
                }
                {
                \Lambda-\frac{1}{2}(\eta^2+\kappa^2)
                }
        +
        \frac{3}{16}\fracbrac{
                \rho^2+\sigma^2
                }
                {
                \Lambda-\frac{1}{2}(\eta^2+\kappa^2)
                }^2
       \right]~.
\end{eqnarray}

\citet{NobiliRoxburgh1986} describe how the apsidal precession caused by
general relativity (GR) can be approximated by including a potential term equal
to 
\begin{eqnarray}
     H_\mathrm{GR}(\eta,\kappa) = -\int_{-\pi}^{\pi}\frac{3G^2M_*^2}{c^2r^2}d\lambda = -\frac{3G^2M_*^2}{c^2a^2\sqrt{1-e^2}}
  =- \frac{3 G^{4} M^{4} \mu^4}
     {\Lambda^{3} c^{2} \left(\Lambda- \frac{1}{2}\left(\eta^{2} + \kappa^2\right)\right)}
     \label{eq:gr_potential}~.
\end{eqnarray}
Primary oblateness and GR effects can be modeled by adding terms in the form
of Equations \eqref{eq:j2_potential} and \eqref{eq:gr_potential} to the
Hamiltonian represented by a \textsf{PoincareHamiltonian} object, described
below in Section \ref{sec:hamiltonian:poincare}, using the
\textsf{add\_orbit\_average\_J2\_terms} and \textsf{add\_gr\_potential\_terms}
methods.
 
\section{Hamiltonian and Poincare Modules}
\label{sec:hamiltonian}
This section begins in Section \ref{sec:hamiltonian:general} with a description
of how the \celmech~code represents general Hamiltonian systems. In Section
\ref{sec:hamiltonian:poincare}, we describe the \textsf{PoincareHamiltonian}
class, which is capable of representing the Hamiltonian of a planetary system
by the inclusion of a collection of disturbing function terms capturing
inter-planet gravitational interactions.  Section
\ref{sec:hamiltonian:canonical_transformations} describes how canonical
transformations, which can be applied to modify any Hamiltonian represented by
\celmech, are implemented.  Section
\ref{sec:hamiltonian:canonical_transformations:lie-series} describes the
implementation of the \textsf{FirstOrderGeneratingFunction} class that can be
used to apply techniques from canonical perturbation theory to ``average" away
fast oscillations in the instantaneous, ``osculating" orbital elements and
transform to ``mean" elements that better track the system's long-term
evolution.  

\subsection{Representing Hamiltonian Systems}\label{sec:hamiltonian:general}

The \textsf{celmech.hamiltonian} module defines the \textsf{PhaseSpaceState}
and \textsf{Hamiltonian} classes which serve as the basic objects used by
\celmech~to represent Hamiltonian dynamical systems.  The Jupyter example
notebook
\href{https://github.com/shadden/celmech/blob/master/jupyter_examples/SimplePendulum.ipynb}{\textsf{SimplePendulum.ipynb}}
provides a basic example that applies these classes to model the motion of a
simple pendulum.  \textsf{PhaseSpaceState} instances represent points in the
phase space of a Hamiltonian dynamical system and store symbolic variables
along with numerical values for canonical variables as key-value pairs of an
ordered dictionary under the \textsf{qp} attribute.  The \textsf{Hamiltonian}
class can be used to define a Hamiltonian function of the phase space
coordinates and momenta of a particular \textsf{PhaseSpaceState} instance.  A
\textsf{Hamiltonian} instance is initialized by passing a symbolic expression
for the Hamiltonian along with a \textsf{PhaseSpaceState} instance.  The
\textsf{PhaseSpaceState} will then be stored by the new \textsf{Hamiltonian}
object as its \textsf{state} attribute.  The symbolic expression for the
Hamiltonian should depend on the same canonical variables stored by the
\textsf{PhaseSpaceState} instance.  The Hamiltonian represented by a
\textsf{Hamiltonian} class can also depend on any number of symbolic parameters
whose numerical values should be set at initialization by passing a dictionary
that pairs parameter symbols with their numerical values.  Numerical parameter
values are stored as the \textsf{H\_params} attribute and can be updated at any
point after initializing a \textsf{Hamiltonian} instance to explore how a
system's dynamics depends on parameter values.

In addition to storing a symbolic representation of the Hamiltonian function
and its associated flow, a \textsf{Hamiltonian} object can be used to integrate
the equations of motion and evolve a phase space trajectory in time.  In order
to do so, a \textsf{Hamiltonian} instance implements functions
\textsf{flow\_func} and \textsf{jacobian\_func} that take numerical values of
the canonical variables as input and return numerical value of the flow given
by Hamilton's equations, and its Jacobian with respect to the canonical
variables.  These function are used in conjunction with the \textsf{scipy}
package's \citep{SciPy} \textsf{integrate.ode} class\footnote{See
\href{https://docs.scipy.org/doc/scipy/reference/generated/scipy.integrate.ode.html}{docs.scipy.org/doc/scipy/reference/generated/scipy.integrate.ode.html}}
to evolve the phase space state of the system.

\subsection{Constructing Hamiltonians for Planetary Systems}
\label{sec:hamiltonian:poincare}

The \textsf{celmech.poincare} module provides tools to model planetary systems'
dynamics by building a Hamiltonian from the disturbing function expansion
described in Section \ref{sec:disturbing_function}.  To build these Hamiltonian
models, the module defines the \textsf{Poincare} class, a special subclass of
the  \textsf{PhaseSpaceState} class, and \textsf{PoincareHamiltonian}, a
special subclass of the \textsf{Hamiltonian} class.  The \textsf{Poincare}
class represents the phase space state of an $N$-body planetary system in terms
of the canonically conjugate coordinates $\pmb{q} =
(\lambda_1,\eta_1,\rho_1,...,\lambda_{N-1},\eta_{N-1},\rho_{N-1})$ and momenta
$\pmb{p} =
(\Lambda_1,\kappa_1,\sigma_1,...,\Lambda_{N-1},\kappa_{N-1},\sigma_{N-1})$.  In
addition to storing these canonical variables, a \textsf{Poincare} instance
owns a collection of \textsf{PoincareParticle} objects through the
\textsf{Poicare.particles} attribute.  The individual members of this
collection represent the individual bodies of the $N$-body system, i.e., the
star and planets.  Each planet \textsf{PoincareParticle} object  records the
values of its mass and six canonical variables as well as various quantities,
including orbital elements, derived from these values. The API for accessing
the mass and orbital element values of individual \textsf{PoincareParticle}
instances is designed to closely mirror that of the \textsf{REBOUND}
\textsf{particle} class. 

A \textsf{PoincareHamiltonian} is initialized by passing a \textsf{Poincare}
object instance, which is then stored as the \textsf{PoincareHamiltonian}'s
\textsf{state} attribute.  Upon initialization, the symbolic Hamiltonian stored
by a \textsf{PoincareHamiltonian} object for an $N$-body planetary system is
simply the sum of $N-1$ individual Keplerian Hamiltonians.  In other words, the
Hamiltonian is given by $H = \sum_{i=1}^{N-1} H_{\mathrm{Kep},i}$ where
$H_{\mathrm{Kep},i}$ is given by Equation \eqref{eq:hamiltonian:Hkep_def}.  The
user can then add specific interaction terms between pairs of planets from the
disturbing function expansion described in Section
\ref{sec:disturbing_function} using a variety of methods of the
\textsf{PoincareHamiltonian} class.  The most basic method for adding
disturbing function terms is \textsf{PoincareHamiltonian.add\_cosine\_term}.
This method takes the integers $\mathbf{k}=(k_1,...,k_6)$ along with inner and
outer planet indices, $i$ and $j$, as input and adds the terms
 \begin{eqnarray}
      -\frac{Gm_im_j}{a_{j,0}}\sum_{\pmb{l},\pmb{\nu}}C_{\bf{k}}^{\pmb{\nu},\pmb{l}}(\alpha_{i,j})
      |Y_i|^{|k_5|+2\nu_1}
      |Y_j|^{|k_6|+2\nu_2}
      |X_i|^{|k_3|+2\nu_3}
      |X_j|^{|k_4|+2\nu_4}
      \delta_i^{l_i}\delta_j^{l_j}
      \cos[{\bf k}\cdot\pmb{\theta}_{ij}]~
      \label{eq:poincare:a_term}
 \end{eqnarray}
to the Hamiltonian.\footnote{
The reference semi-major axes $a_{i0}$ occurring Equation
\eqref{eq:poincare:a_term}, both explicitly as well as implicitly via the
variables $\alpha_{ij}$ and $\delta_i$, are stored as parameters of the
\textsf{PoincareHamiltonian} object as parameters in the \textsf{H\_params}
attribute.  The reference semi-major axis values are set to the instantaneous
semi-major axes of the particles at the time when the
\textsf{PoincareHamiltonian} is initialized.  Users should be aware that
altering the values of the $a_{i0}$ parameters stored by a
\textsf{PoincareHamiltonian} object will \emph{not} automatically update the
values of the $\Lambda_{i,0}$ or the disturbing function coefficient parameters
stored by a \textsf{PoincareHamiltonian} object.  The simplest way to update
all parameters depending on the reference semi-major axes in a self-consistent
fashion is initialize a new \textsf{PoincareHamiltonian} object from a set of
particles possessing instantaneous semi-major axes equal to the desired
reference values.  
} 
The combinations of $\pmb{l}$ and $\pmb{\nu}$ occurring in the sum are
controlled by the user through a variety of optional keyword arguments. If no
keyword arguments are passed, the default behavior is to include only the
lowest order contribution from the disturbing function, i.e., a single term
with $\pmb{l}=(0,0)$ and $\pmb{\nu} = (0,0,0,0)$.  The
\textsf{PoincareHamiltonian} methods \textsf{add\_MMR\_terms} and
\textsf{add\_secular\_terms} combine this basic \textsf{add\_cosine\_term}
method with the functions for enumerating particular disturbing function
arguments described in Section \ref{sec:disturbing_function:arguments} to add
collections of resonant or secular terms to the Hamiltonian up to a
user-specified order in inclinations and eccentricities.

\subsection{Canonical Transformations}
\label{sec:hamiltonian:canonical_transformations}

The study of Hamiltonian systems is often greatly simplified by canonical
transformations from one set of canonical variables to another that leave
Hamilton's equations unchanged.  This allows for the algorithmic determination
of constants of motion that can sometimes be used to greatly reduce the
dimensionality of the problem \citep[e.g.,][]{Hadden2019}.  The
\celmech~\textsf{CanonicalTransformation} class provides a general framework
for implementing canonical transformations. Instances of this class allow the
user to apply canonical transformations to expressions by applying substitution
rules that replace any occurrences old canonical variables with new ones and
vice versa.  A \textsf{CanonicalTransformation} instance can also be used to
generate new \textsf{Hamiltonian} objects by applying its transformation to an
existing \textsf{Hamiltonian} instance's expression for the Hamiltonian, stored
as the \textsf{H} attribute. The resulting \textsf{Hamiltonian} object's
\textsf{state} attribute will be a new \textsf{PhaseSpaceState} object
representing the new canonical variables and its \textsf{H} attribute will be
the transformed Hamiltonian.

Users can create canonical transformation objects by supplying sets of old and
new canonical variables along with substitution rules.\footnote{ It is the
user's responsibility to ensure that the supplied rules actually constitute a
canonical change of variables, but the method \textsf{test\_canonical()} can be
used to test whether the supplied rules constitute a canonical transformation.}
User's can also use one the following class methods for initializing some
frequently-used canonical transformations:

\begin{itemize} \item 
					\textsf{from\_type2\_generating\_function} allows the
					user to specify a generating function
					$F_2(\pmb{q},\pmb{P})$ producing a canonical
					transformation that satisfies the equations
					\begin{eqnarray} \pmb{Q} = \nabla_{\pmb{P}}F_2~&;~
					\pmb{p} = \nabla_{\pmb{q}}F_2~.  \end{eqnarray}
					Provided an expression for the generating function,
					$F_2$,  this method will attempt to automatically
					determine the corresponding transformation rules by
					applying \textsf{sympy}'s \textsf{solve} function to
					express the new variables $(\pmb{Q},\pmb{P})$ in terms
					of the old variables  $(\pmb{q},\pmb{p})$ and vice
					versa. 
				\item 
					\textsf{cartesian\_to\_polar} produces a
					canonical transformation that transforms
					user-specified conjugate coordinate-momentum
					pairs of old variables $(q_i,p_i)$ to new
					conjugate pairs
					$(Q_i,P_i)=\left((\tan^{-1}(q_i/p_i),\frac{1}{2}(q_i^2+p_i^2)\right)$.
					All other conjugate pairs will remain unchanged.  
				\item 
					\textsf{polar\_to\_cartesian} produces a
					canonical transformation that transforms
					user-specified conjugate pairs, $(q,_i,p_i)$,
					to new conjugate coordinate-momentum pairs
					$(Q_i,P_i)=\left(\sqrt{2p_i}\sin
					q_i,\sqrt{2p_i}\cos q_i\right)$.  All other
						conjugate pairs will remain unchanged.  
				\item
					\textsf{from\_linear\_angle\_transformation}
					produces the canonical transformation
					$(\pmb{q},\pmb{p})\rightarrow(\pmb{Q},\pmb{P})$
					given by \begin{eqnarray} \pmb{Q} = T \cdot
					\pmb{q} ~&;~ \pmb{P} = (T^{-1})^\mathrm{T}
					\cdot \pmb{p} \end{eqnarray} where $T$ is a
					user-supplied, invertible matrix. 
				\item
					\textsf{from\_poincare\_angles\_matrix} takes a
					\textsf{Poincare} instance as input, along with
					an invertible matrix $T$, and produces a
					transformation where the new canonical
					coordinates are linear combinations of the
					planets' angular orbital elements given by
					$\pmb{Q} = T\cdot (\lambda_1 ,..., \lambda_N,
					-\varpi_1,... ,-\varpi_N -\Omega_1 ,...,
					-\Omega_N) $ and the corresponding new
					conjugate momenta are given in terms of the
					canonical modified Delaunay variables described
					in Section \ref{sec:canonical_coordinates} by
					$\pmb{P} = (T^{-1})^\mathrm{T}\cdot
				(\Lambda_1,...,\Lambda_N,\Gamma_1,...,\Gamma_N,Q_1,...,Q_N)$.
				\item 
					\textsf{Lambdas\_to\_delta\_Lambdas} takes a \textsf{Poincare}
					instance as input and implements the canonical transformation
					replacing the $\Lambda_i$ variables with $\delta\Lambda_i =
					\Lambda_i-\Lambda_{i,0}$ as the  momenta canonically conjugate
					to planets' mean longitudes, $\lambda_i$.  
				\item
					\textsf{rescale\_transformation} produces a canonical
					transformation that simultaneously rescales the
					Hamiltonian and all canonical momentum variables by a
					common factor, $f$. In other words, if $H$ is the old
					Hamiltonian and  $p_i$ are the old momentum variables,
					the new Hamiltonian will be $H' = fH$ and $p_i' = fp_i$
					will replace $p_i$ as momentum variables conjugate to
					the old coordinates $q_i$. The user can optionally
					specify a subset of canonical variable pairs
					$(y_i,x_i)$ as Cartesian pairs which will be
					transformed so that the new, re-scaled variables are
					instead $(y'_i,x'_i) = (\sqrt{f}y_i,\sqrt{f}x_i)$.	
				\item 
					\textsf{composite} combines a series of canonical transformations.
\end{itemize}
While time-dependent canonical transformations are not currently supported,
they can always be implemented by re-formulating the Hamiltonian under
consideration in the extended phase space that includes time and the negative
of the system's energy as an additional conjugate variable pair.

Often a canonical transformations are applied to eliminate the explicit
dependence of the transformed Hamiltonian on one or more canonical variables.
This is because, when the transformed Hamiltonian does not depend on one of the
angle variables, its the corresponding conjugate momentum variable is conserved
and Hamilton's equations for the remaining degrees of freedom are simplified.
To make the discussion more concrete, suppose we have a canonical
transformation of the canonical variables of an $N$ degree of freedom system
$(q_i,p_i)\rightarrow
(\phi_1,...,\phi_M,Q_1,....,Q_{N-M},I_1,...,I_M,P_1,...,P_{N-M})$ where the
transformed Hamiltonian, $H'(Q_i,P_i;I_j)$, is independent of the coordinate
variables $\phi_j$. so that the quantities $I_1,...,I_M$ are constant.  When
such an elimination can be accomplished, it is frequently convenient to
consider a system's dynamics in the reduced phase space comprised of the
remaining $N-M$ degrees of freedom with canonical variables $(Q_i,P_i)$ and
treat the $M$ conserved quantities, $I_j$, as parameters of the Hamiltonian.
When transforming a Hamiltonian by applying the
\textsf{old\_to\_new\_hamiltonian} method of a \textsf{CanonicalTransformation}
object, the user has the option to automatically detect whether the resulting
transformed Hamiltonian is independent of any of the new canonical variables
and, if so, produce a \textsf{Hamiltonian} object with a \textsf{state}
attribute respresenting a point in the reduced phase space with canonical
variables $(Q_i,P_i)$ and with the quantities $I_j$ stored as parameters under
the \textsf{Hamiltonian} object's \textsf{H\_params} attribute.  To enable the
application of the inverse transformation from the new canonical variables back
to old ones in this situation, the new \textsf{Hamiltonian} object possess a
\textsf{full\_qp} attribute that stores the full set of variables,
$(\phi_1,...,\phi_M,Q_1,....,Q_{N-M},I_1,...,I_M,P_1,...,P_{N-M})$, while the
\textsf{qp} attribute\footnote{The \textsf{qp} attribute of a
\textsf{Hamiltonian} object is actually just a shortcut for
\textsf{Hamiltonian.state.qp} where  \textsf{Hamiltonian.state} is the
\textsf{PhaseSpaceState} object representing the state of the system. See
Section \ref{sec:hamiltonian:general}.} only stores the variables of the
reduced phase space, $(Q_i,P_i)$.  The reduced \textsf{Hamiltonian} object's
\textsf{full\_qp} attribute will only store the initial values of the ignorable
coordinates, $\phi_j$, determined at the time the transformation was applied.
It is important to recognize the the actual time evolution of these
coordinates, given by  $\frac{d}{dt}\phi_j=\frac{\partial}{\partial
I_j}H'(Q_i,P_i;I_j)$, is generally non-trivial and depends on the specific
trajectory, $(Q_i(t),P_i(t))$, of the system in the reduced phase space.
Consequently, it will usually not be possible to re-construct the full phase
space state of the system in the old canonical variables $(q_i,p_i)$ if only
the reduced equations of motion governing the $(Q_i,P_i)$ are integrated.
Nonetheless, in many applications the values of these ignorable coordinates are
unimportant for the aspects the dynamics one wishes to study and application of
the inverse transformation to the data stored in the \textsf{full\_qp}
attribute can provide useful information about how the system evolves in the
original canonical variables.  A basic illustration of these principles is
provided in an
\href{https://github.com/shadden/celmech/blob/master/jupyter_examples/BasicCanonicalTransformation.ipynb}{example
Jupyter notebook}.

\subsection{Lie series transformations}
\label{sec:hamiltonian:canonical_transformations:lie-series}

Modeling the dynamics of a planetary system with a Hamiltonian that includes
only a finite number of terms from a disturbing function can be justified
because the infinite number of ignored terms are rapidly oscillating and have
negligible average influence on the motion, at least over sufficiently short
timescales.  Nevertheless, these fast oscillations in the full problem can make
it challenging to match initial conditions in an ``averaged" formulation, and
such discrepancies can make comparisons between analytical and $N$-body models
challenging, especially near regions of phase space where the dynamical
behavior changes sharply, e.g., near resonance boundaries.  Additionally, while
typically we are interested in methods for {\it removing} these nuisance terms
in the disturbing function, there are also cases where these fast oscillations
are of primary interest.  An important example is for transiting pairs of
planets near (but not in) mean motion resonances, where the ``fast"
oscillations due to the resonant terms are observable as transit timing
variations \citep[for a Lie sires approach to TTVs, see][]{Nesvorny2008}.

A powerful approach for transforming back and forth between instantaneous, or
``osculating", canonical variables (which are subject to a multitude of fast
oscillations) and ``mean" variables that remove these fast modes, is through
the Lie series method.  This technique introduces a near-identity canonical
transformation (ensuring that the ``mean" variables remain canonical) that, by
construction, can cancel any set of rapidly oscillating terms in the
Hamiltonian to first order in the planet-star mass ratio.  This transformation
is often referred to as ``averaging" since, at this leading order in the
masses, the technique is equivalent to averaging the problem over the rapidly
oscillating angles.  An advantage of the Lie series method is that it
explicitly constructs the resultant terms at higher orders in the masses, which
are not available through an averaging approach.  A thorough discussion of
canonical perturbation theory using the Lie series method, including important
issues related to chaos, non-integrability, and the (non-)convergence of these
series, is beyond the scope of this paper and can be found in references such
as \citet{Morbidelli2002} or \citet{lichtenberg2013}.  We will limit our
discussion to the construction of canonical transformations to first order in
the planet-star mass ratios, which are equivalent to transformations back and
forth between osculating and mean variables, and are implemented in
\textsf{celmech}. 

To introduce Lie-series generating functions, consider splitting the
Hamiltonian of a $N$-body system  into three pieces: a Keplerian piece given by
$H_\mathrm{Kep} =\sum_{i=1}^{N-1}H_{\mathrm{Kep},i}$, a term $H_\mathrm{slow}$
contains slowly varying disturbing function terms such as those with (near)
resonant cosine arguments or secular terms, and a term  $H_\mathrm{osc}$ that
contains rapidly oscillating with negligible  influence on the dynamics.  We
seek a canonical transformation that eliminates $H_\mathrm{osc}$ to first order
in planet masses.  In other words, we seek a canonical transformation $T: ({\bf
q },{\bf p}) \rightarrow ({\bf q' },{\bf p'})$ such that the transformed
Hamiltonian is $H' \equiv H \circ T^{-1}=H_{\mathrm{Kep}} + H_\mathrm{slow} +
\mathcal{O}((m/M_*)^2)$.  We use the Lie series method to construct such a
transformation.  The Lie transformation, $\exp[\mathcal{L}_\chi]$, of a
function of phase space coordinates, $f$, is defined in terms of the Lie
derivative operator, $\mathcal{L}_\chi\equiv [\cdot,\chi]$, as
\begin{eqnarray}
     \exp[\mathcal{L}_\chi]f =\sum_{n=0}^\infty \frac{1}{n!}\mathcal{L}_\chi^{n}f=f + [f,\chi] + \frac{1}{2}[[f,\chi],\chi]+...~.
\end{eqnarray}
Since a Lie transformation evolves $f$ under the flow induced by Hamilton's
equations for the ``Hamiltonian" $\chi$ it is canonical by construction.  The
Lie series method seeks to find a generating function $\chi$ such that the Lie
transformation $\exp[\mathcal{L}_\chi]$ produces the desired transformation,
$T$, that eliminates the unwanted terms.  If the generating function, $\chi$,
is itself of order $\mathcal{O}(m/M_*)$ then the transformed Hamiltonian is $H'
= \exp[\mathcal{L}_{\chi}]H \approx H_{\mathrm{Kep}} + H_\mathrm{slow} +
H_\mathrm{osc} + [H_{\mathrm{Kep}},\chi] $ after dropping terms $
\mathcal{O}((m/M_*)^2)$ and the generating function $\chi$ should satisfy
$[H_{\mathrm{Kep}},\chi] = -H_\mathrm{osc}$.  Noting that the Poisson bracket
of $H_{\mathrm{Kep}}$ with all canonical variables except the mean longitudes
vanishes and that 
\begin{eqnarray}
    \left[ H_{\mathrm{Kep}} , \frac{\sin(\mathbf{k}\cdot{\pmb{\theta}_{i,j}})}{k_1n_j + k_2 n_i}\right] = -\cos(\mathbf{k}\cdot{\pmb{\theta}_{i,j}})
\end{eqnarray}
where $n_i =\partial H_{\mathrm{Kep}} / \partial \Lambda_i$ is the $i$th
planet's mean motion, we immediately see that any disturbing function terms in
$H_\mathrm{osc}$ involving planets $i$ and $j$ written in the form of Equation
\eqref{eq:poincare:a_term} will be eliminated to first order in the planet
masses by simply including a corresponding term in $\chi$ that replaces
$\cos(\mathbf{k}\cdot{\pmb{\theta}_{i,j}})$ with
${\sin(\mathbf{k}\cdot{\pmb{\theta}_{i,j}})}/({k_1n_j + k_2 n_i})$ provided
$k_1n_j + k_2 n_i\ne 0$.  The condition that $k_1n_j + k_2 n_i\ne 0$ should
always be satisfied since $H_\mathrm{osc}$ is presumed to be comprised only of
rapidly oscillating terms.

The \celmech~class \textsf{FirstOrderGeneratingFunction} allows users to
construct a generating function $\chi$ that eliminates a user-specified
collection of disturbing function terms in the transformed Hamiltonian to first
order in planet masses following the approach just described above.  Terms are
added to the generating function in much the same way that terms are added to a
\textsf{PoincareHamiltonian} instance. In fact,
\textsf{FirstOrderGeneratingFunction} is a subclass of the
\textsf{PoincareHamiltonian} of \textsf{PoincareHamiltonian} that overwrites
the parent class's routines for adding cosine terms so that the trigonometric
terms $\cos(\mathbf{k}\cdot{\pmb{\theta}_{i,j}})$ are instead replaced with
${\sin(\mathbf{k}\cdot{\pmb{\theta}_{i,j}})}/({k_1n_j + k_2 n_i})$.  The class
provides a symbolic representations of the generating function under the
$\textsf{chi}$ and $\textsf{N\_chi}$ attributes, similar to the
\textsf{PoincareHamiltonian} class's \textsf{H} and \textsf{N\_H} attributes.
The class also provides an array of methods for transforming canonical
variables and deriving perturbative solutions to the equations of motion. These
are demonstrated in example notebooks
\href{https://github.com/shadden/celmech/blob/master/jupyter_examples/FirstOrderGeneratingFunction.ipynb}{here}
and
\href{https://github.com/shadden/celmech/blob/master/jupyter_examples/TransitTimingVariations.ipynb}{here}. 

In addition to term-wise elimination of individual rapidly oscillating
disturbing function terms described above, It is possible to remove all
harmonics depending solely on the angular combination $\psi =
\lambda_i-\lambda_j$ to zeroth order in eccentricities and inclinations with a
single transformation.  A number of authors  have previously exploited the fact
that perturbative solutions for planets' mutual perturbation at zeroth order in
eccentricities and inclinations can be expressed in closed form without needing
to resort to an expansion in Fourier harmonics
\citep[e.g.,][]{Malhotra1993,Agol2005,Nesvorny2014,HaddenTESS2019}, though not
all these studies deploy a Lie series formalism.  To zeroth order in
inclinations and eccentricities, the disturbing function, $ {\cal
R}^{(i,j)}_\mathrm{dir.} + {\cal R}^{(i,j)}_\mathrm{ind.}$, given in  Equation
\eqref{eq:disturbing_function:df_dir_ind} can be written as
\begin{eqnarray}
 {\cal R}^{(i,j)}_\mathrm{dir.} + {\cal R}^{(i,j)}_\mathrm{ind.}
 \approx 
  P(\lambda_j-\lambda_i;\alpha) - \frac{1}{\sqrt{\alpha}}\cos(\lambda_i-\lambda_j)
  +\mathcal{O}(e,I^2)
  \label{eq:df_0th_order}
\end{eqnarray}
where $P(\psi;\alpha)=(1 + \alpha^2 - 2\alpha\cos(\lambda_j -
\lambda_i))^{-1/2}$.  Therefore, a Lie series transformation with generating
function
\begin{eqnarray}
        \chi &=& -\frac{Gm_1m_2}{a_2(n_1-n_2)}\left(-\frac{1}{\sqrt{\alpha}}\sin({\lambda_i-\lambda_j}) + \int^{\lambda_i-\lambda_j}_{0}\left(P(\psi,\alpha) - \bar{P}(\alpha)\right) d\psi \right)~,
        \label{eq:chi_0th_order}
\end{eqnarray}
where $\bar{P}(\alpha) = \frac{1}{2\pi}\int_{-\pi}^{\pi}P(\psi,\alpha)d\psi$ is
the mean value of $P(\psi,\alpha)$, will eliminate the disturbing function
terms in Equation \eqref{eq:df_0th_order}.  The integral in Equation
\eqref{eq:chi_0th_order} can be written explicitly in terms of elliptic
integrals as 
\begin{eqnarray}
    \int^{\psi}_{0}\left(P(\psi',\alpha) - \bar{P}(\alpha)\right) d\psi' 
    =  
    \frac{2}{1-\alpha}F\left(\frac{\psi}{2}\bigg|-\frac{4\alpha}{(1-\alpha)^2}\right)
     - 
     \frac{2}{\pi}\mathbb{K}(\alpha^2)\psi
\end{eqnarray}
where $\mathbb{K}$ is the complete elliptic integral of the first kind and
$F(\cdot | m)$  is an incomplete elliptic integrals of the first kind with
modulus $m$.  A term of the form given in Eqauation \eqref{eq:chi_0th_order}
can be added to the symbolic generating function stored by a
\textsf{FirstOrderGeneratingFunction} object with the
\textsf{add\_zeroth\_order\_term} method.

\section{Summary}

In this paper we presented \celmech, an open-source \textsf{python} package for
conducting celestial mechanics calculations. The code combines a disturbing
function expansion algorithm based on a method originally devised
\citet{EllisMurray2000} with the symbolic mathematics capabilities provided by
the \textsf{sympy} library \citep{sympy} to enable users to efficiently create
and integrate Hamiltonian models for planetary systems. The \celmech~API is
designed so that users can easily compare these Hamiltonian models to direct
$N$-body integrations with the \rebound~code \citep{ReinLiu2012}.

The code is available through the \href{https://pypi.org/}{Python Package Index
(PyPI)} as well as on GitHub at
\href{https://github.com/shadden/celmech}{github.com/shadden/celmech}.  The
code can can be freely redistributed under the {GPL-3.0 license}.  Online
documentation can be found at
\href{https://celmech.readthedocs.io/en/latest/}{celmech.readthedocs.io}.  A
library of Jupyter notebook examples illustrating various features and
applications of the code is available at
\href{https://github.com/shadden/celmech/tree/master/jupyter_examples}{github.com/shadden/celmech/tree/master/jupyter\_examples}.

The \celmech~code's modular design will allow ongoing development.  Some
features currently implemented in \celmech, in addition to the modules and
capabilities detailed in this paper, include tools for formulating and
integrating equations of motion governing MMRs that are derived from
numerically averaged disturbing functions (rather than truncated power series
expansions in eccentricity and inclination) and tools for formulating and
integrating secular equations of motion. We plan to detail additional features
in forthcoming papers. 

\vspace{5mm}
\software{
\textsf{matplotlib} \citep{Matplotlib}, 
\textsf{NumPy} \citep{numpy},
\textsf{REBOUND} \citep{ReinLiu2012}, 
\textsf{REBOUNDx} \citep{reboundx}, 
\textsf{scipy} \citep{scipy_all},
\textsf{sympy} \citep{sympy}
}

\acknowledgments We thank David Hernandez for helpful discussions.
Implementations of some \celmech~routines are based on the Mathematica package
by Fabio Zugno available at
\href{https://library.wolfram.com/infocenter/MathSource/4256/}{library.wolfram.com/infocenter/MathSource/4256/}.
S.H. acknowledges support by the Natural Sciences and Engineering Research
Council of Canada (NSERC), funding reference \#CITA 490888-16.  D.T. is
grateful for support from the Lyman Spitzer Jr. fellowship.

\appendix
\section{Table of Symbols}
\label{app:list_of_symbols}

Table \ref{tab:symbols} lists the main mathematical symbols used in the text
along with brief descriptions as appropriate.

\begin{table*}\label{tab:symbols}
\caption{ Summary of mathematical notation used throughout the paper.  We
		include short definitions and/or references to equations where
		quantities are defined when appropriate.  }
\centering
\begin{tabular}{| c | c | l |}
  \hline
  Name & Expression & Description \\
  \hline
  $G$ & & Newton's gravitational constant \\
  $M_*$ & & Mass of central star \\
  $m_i$ & & Mass of $i$th planet \\
  $\pmb{r}_i$ & & Heliocentric position of $i$th planet  \\
  $\tilde{\pmb{r}}_i$ & & Barycentric momentum of $i$th planet  \\
  $a_i$ & & Semi-major axis of $i$th planet\\
  $e_i$ & & Eccentricity of $i$th planet \\
  $I_i$ & & Inclination of $i$th planet \\
  $\lambda_i$ & & Mean longitude of $i$th planet \\
  $\varpi_i$ & & Longitude of periapse $i$th planet \\
  $\Omega_i$ & & Longitude of ascending node $i$th planet \\  
  $s_i$ & $\sin(I_i/2)$& Inclination variable appearing in many classic disturbing function expansions \\
  $\mu_i$ &  $m_iM_*/(M_*+m_i)$ & Reduced mass of $i$th planet \\
  $M_i$ &  $M_* + m_i$ & Effective central mass for $i$th planet's orbit \\
  $\Lambda_i $ & $\mu_i\sqrt{GM_ia_i}$ & Canonical momentum conjugate to $\lambda_i$ \\
  $\Gamma_i $ & $ \Lambda_i(1-\sqrt{1-e_i^2})$ & Canonical momentum conjugate to $\gamma_i=-\varpi_i$\\
  $Q_i $ & $ = 2\Lambda_i\sqrt{1-e_i^2}\sin^2(I_i/2)$& Canonical momentum conjugate to $q_i=-\Omega_i$\\
  $\kappa_i$ & $= \sqrt{2\Gamma_i}\cos{ \varpi_i}$&\\
  $\eta_i$ & $= -\sqrt{2\Gamma_i}\sin{ \varpi_i}$&\\  
  $\sigma_i$ & $= \sqrt{2Q_i}\cos{ \Omega_i}$&\\
  $\rho_i$ & $= -\sqrt{2Q_i}\sin{ \Omega_i}$&\\  
  $\mathcal{R}^{(i,j)}_\mathrm{dir.}$&$=\frac{a_j}{|\mathbf{r}_j- \mathbf{r}_i|}$& Direct component of disturbing function\\
  $\mathcal{R}^{(i,j)}_\mathrm{ind.}$&$=\frac{a_j}{GM_*m_jm_i}\tilde{\mathbf{r}}_j \cdot \tilde{\mathbf{r}}_i$& Indirect component of disturbing function\\
  $\alpha_{i,j}$ & $=a_i/a_j$& Semi-major axis ratio\\  
  $\pmb{\theta}_{i,j}$ & $=(\lambda_j,\lambda_i,\varpi_i,\varpi_j,\Omega_i,\Omega_j)$& Vector of angle variables \\  
  $\tilde{C}_{\bf k}^{\pmb{\nu}}(\alpha)$ & & Disturbing function coefficient of $\cos(\pmb{k}\cdot\pmb{\theta}_{i,j})$ for expansion in $e_i$ and $s_i$  \\
  $b_s^{(s)}(\alpha)$&$\frac{1}{\pi}\int_{-\pi}^{\pi}\frac{\cos(j\theta)}{(1+\alpha^2-2\alpha\cos\theta)^{s}}$& Laplace coefficient\\
  $z_i$ & $=e_i\exp(\mathrm{i}\varpi_i) $ & Complex eccentricity \\
  $\zeta_i$ & $=s_i\exp(\mathrm{i}\Omega_i) $ & Complex inclination variable \\
  $\Lambda_{i,0}$ && Reference value for $\Lambda_i$\\ 
  $\delta_i$ & $\Lambda_i/\Lambda_{i,0} - 1$& Fractional deviation of $\Lambda_i$ from reference value $\Lambda_{i,0}$  \\
  $X_i$ & $=\sqrt{\frac{1}{\Lambda_{i,0}}}(\kappa_i-\mathrm{i}\eta_i)$ & $\approx e_i\exp(\mathrm{i}\varpi_i) + \mathcal{O}(e_i^3)$ \\
  $Y_i$ & $=\sqrt{\frac{1}{4\Lambda_{i,0}}}(\sigma_i - \mathrm{i}\rho_i)$ & $\approx \sin(I_i/2)\exp(\mathrm{i}\Omega_i) + \mathcal{O}(I_ie_i^2)$ \\
  ${C}_{\bf k,\pmb{l}}^{\pmb{\nu}}(\alpha)$ & & Disturbing function coefficient of $\cos(\pmb{k}\cdot\pmb{\theta}_{i,j})$ for expansion in $X_i$,$Y_i$, and $\delta_i$  \\  
  \hline
\end{tabular}
\end{table*}
\clearpage

\section{Expansion of the Indirect Disturbing Function}
\label{app:indirect}

In this appendix we detail the expansion of the indirect component of the
disturbing function, $${\cal R}^{(i,j)}_\mathrm{ind.} =
-\frac{a_j}{GM_*m_im_j}\tilde{\pmb{r}}_i \cdot \tilde{\pmb{r}}_j~,$$ closely
following \citet{Laskar1995}, who present a method for deriving a complete
expansion of the indirect disturbing function, best accomplished with the aid
of computer algebra. We extend their derivation in order to provide explicit
formulas for disturbing function coefficients associated with specific cosine
arguments.  Writing $\tilde{\pmb{r}}_i = \mu_i n_ia_i{\pmb v}_i$, the indirect
disturbing function can be written as
\begin{equation}
    {\cal R}^{(i,j)}_\mathrm{ind.} 
    = 
        -\frac{n_i n_j a_j^2a_i}{GM_*}{\pmb v}_i \cdot {\pmb v}_j
    \approx 
        -\frac{1}{\sqrt{\alpha}}{\pmb v}_i \cdot {\pmb v}_j + {\cal O}((m/M_*)^2)~.
\end{equation}
where the ${\cal O}((m/M_*)^2)$ error terms arise from approximating
$\mu_i\approx m_i$.  The vectors ${\pmb v}_i$ may be expressed in terms of
orbital elements as
\begin{eqnarray}
    {\pmb v}_i &=& \frac{1}{\sqrt{1-e_i^2}}{\cal R}_3(\Omega_i){\cal R}_1(I_i){\cal R}_3(-\Omega_i) \cdot \begin{pmatrix}
    \Re\bparen{\ii \exp[\ii \varpi_i](e^{\ii f_i}+ e_i ) }\\
    \Im\bparen{\ii \exp[\ii \varpi_i](e^{\ii f_i}+ e_i ) }\\
    0
    \end{pmatrix}\\
    &=& \Re\bparen{
    \ii (e^{\ii(f+\varpi)}+e_ie^{\ii\varpi})
    \begin{pmatrix}
    \cos^2(I_i/2) + \bar{\zeta}^2 \\
    -\ii(\cos^2(I_i/2) - \bar{\zeta}^2) \\
    -2\ii \cos(I_i/2)\bar\zeta
    \end{pmatrix}
    }
\end{eqnarray}
where $f_i$ is the planet's true anomaly and $\Re[\cdot]$ and $\Im[\cdot]$
denote the real and imaginary part of the enclosed expression, respectively.
Using the identity $\Re[{A}]\Re[{B}] = \frac{1}{2}\Re\paren{AB + A\bar{B}}$ and
denoting ${\cal Z}_i =\ii (e^{\ii(f+\varpi)}+e_ie^{\ii\varpi})(1-e_i)^{-1/2}$
we have 
\begin{eqnarray}
{\pmb v}_i \cdot {\pmb v}_j    = 
\Re\bparen{
    \underbrace{
        {\cal Z}_i{\cal Z}_j
        \paren{
            \bar{\zeta}_i\cos(I_j/2) -\bar{\zeta}_j\cos(I_i/2)
        }^2
    }_{\text{Term 1}}
 + 
    \underbrace{
        {\cal Z}_i\bar{\cal Z}_j
        \paren{
            \cos(I_j/2)\cos(I_i/2) +
            \bar{\zeta}_i{\zeta}_j
        }^2
    }_{\text{Term 2}}~.
}\label{eq:ui_dot_uj}
\end{eqnarray}
Below we will treat the disturbing function terms arising from the expressions
labeled `Term 1' and `Term 2' in Equation \eqref{eq:ui_dot_uj} separately. But
before doing so, we express ${\cal Z}_i$ explicitly in terms of the mean
longitude, ${\lambda}_i$, rather than the true anomaly, $f_i$, using Hansen
coefficients \citep[e.g.,][]{Hughes1981} to write
\begin{eqnarray}
    e^{\ii f}= \sum_{k=-\infty}^{\infty} X_{k}^{0,1}(e)e^{ik(\lambda-\varpi)}
            = \sum_{k=-\infty}^{\infty}
            \left(
            e^{|k-1|}
            \sum_{\sigma=0}^{\infty}
            N_{k,\sigma}^{0,1}e^{2\sigma}
            \right)
        e^{i(\lambda-\varpi)}~.
\end{eqnarray}
where the coefficients, $N_{k,\sigma}^{0,1}$, appearing in the series expansion
of $X_{k}^{0,1}(e)$ can be calculated with the aid of recursion relations
described in \citet[][]{Hughes1981} and \citet[][]{MD1999ssd}.\footnote{Note
that the coefficients $N_{k,\sigma}^{0,1}$ are \emph{not} equivalent to the
``Newcomb operators" defined in \citet[][]{Hughes1981} and
\citet[][]{MD1999ssd} and denoted as $X^{a,b}_{c,d}$. Rather, the coefficients
$N_{k,\sigma}^{0,1}$  are related to the Newcomb operators according to
$N_{k,\sigma}^{0,1} = X^{0,1}_{k-1 + \sigma,\sigma}$ for $k\ge1$ and
$N_{k,\sigma}^{0,1} = X^{0,1}_{\sigma,1-k+\sigma}$ for $k<1$.  } We define the
related set of coefficients ${\cal X}_{k}^{0,1}(e) =
(1-e^2)^{-1/2}X_{k}^{0,1}(e)$ so that we may write
\begin{eqnarray}
    {\cal Z} = \ii e^{\ii \varpi} \sum_{k\ne 0}{\cal X}_{k}^{0,1}(e)e^{ik(\lambda-\varpi)}
\end{eqnarray}
after noting that no $k=0$ term appears in the sum because $X_{0}^{0,1}(e) =
-e$. We also define coefficients $\mathcal{N}^{0,1}_{k,p} =
\sum_{n=0}^{p}\binom{-1/2}{n}(-1)^n{N}^{0,1}_{k,p-n}$, so that we may write
\begin{eqnarray}
    {\cal X}^{0,1}_{k}(e) &=& e^{|k-1|}\sum_{p=0}^{\infty}{\cal N}^{0,1}_{k,p}e^{2p}~.
\end{eqnarray}
We now turn to deriving expression for the terms labelled `Term 1' and `Term 2'
in Equation \eqref{eq:ui_dot_uj} in terms of orbital elements. 

{\bf Term 1:} Expressing Term 1 in terms of orbital elements, we have
\begin{multline}
 {\cal Z}_i\bar{\cal Z}_j
\paren{
    \cos(I_j/2)\cos(I_i/2) +
    \bar{\zeta}_i{\zeta}_j
    }^2 =\\
e^{\ii(\varpi_i-\varpi_j)}
    \paren{
        (1-s_i^2)(1-s_j^2) 
        +    
        2s_is_j\sqrt{1-s_i^2}\sqrt{1-s_j^2}e^{\ii(\Omega_j - \Omega_i)} 
        +
        s_i^2s_j^2e^{2\mathrm{i}(\Omega_j - \Omega_i)}
    }\\
    \times
    \sum_{k,k'\ne 0}{\cal X}_{k}^{0,1}(e_i){\cal X}_{k'}^{0,1}(e_j)e^{\ii(k\lambda_i- k'\lambda_j - k\varpi_i +k'\varpi_j)}~.
    \label{eq:indirect:term1}
\end{multline}
We see from Equation \eqref{eq:indirect:term1} that Term 1 contributes to the
cosine amplitudes of Equation \eqref{eq:disturbing_function:Hint_fourier} for
terms with cosine arguments satisfying 
\begin{equation}
    \pm\pmb{k}\cdot\pmb{\theta}_{i,j} = -k'\lambda_j + k\lambda_i + (1-k)\varpi_i +  (k'-1)\varpi_j  +  m (\Omega_j - \Omega_i)
\end{equation}
with $m=0,1,2$ and $k,k'\ne0$.
The indirect contributions to the amplitudes will be 
\begin{eqnarray}
    \left[\tilde{C}_{\pmb{k}}^{(\nu_1,\nu_2,\nu_3,\nu_4)}(\alpha)\right]_\mathrm{ind.} 
     &=& 
   -\frac{1}{\sqrt{\alpha}}
   {\cal N}^{0,1}_{k',\nu_4}
   {\cal N}^{0,1}_{k,\nu_3}
   (-1)^{\nu_1+\nu_2}
   \binom{1 - \frac{m}{2}}{\nu_1}
   \binom{1 - \frac{m}{2}}{\nu_2}
   (1 +\delta_{m,1} )
\end{eqnarray}
for the terms with $\mathbf{k}=\pm(k',-k,k-1,1-k',m,-m)$ with $m=0,1$ or 2.  In
particular, indirect contributions to terms associated with an $p$:$p-q$ MMR
will occur for coefficients $\tilde{C}_{(p,q-p,p-q-1,1-p,m,-m)}^{\mathbf{\nu}}$
and $\tilde{C}_{(p,q-p,p-q+1,-1-p,-m,m)}^{\mathbf{\nu}}$ when $(k',k)=(p,p-q)$
and $(k',k)=(-p,q-p)$, respectively.

{\bf Term 2:}  
Writing Term 2  from Equation \eqref{eq:ui_dot_uj} in terms of orbital
elements, we obtain
\begin{multline}
 {\cal Z}_i{\cal Z}_j
\paren{
    \bar{\zeta}_i\cos(I_j/2) -\bar{\zeta}_j\cos(I_i/2)
    }^2 =\\ 
    e^{\ii(\varpi_i+\varpi_j)}\paren{
    s_i^2(1-s_j^2)e^{-2\ii\Omega_i} 
    - 
    2s_is_j\sqrt{(1-s_i^2)(1-s_j^2)}e^{-\ii(\Omega_i+\Omega_j)} 
    + 
    s_j^2(1-s_i^2)e^{-2\ii\Omega_j}
} \\
\times
\sum_{k,k'\ne 0}{\cal X}_{k}^{0,1}(e_i){\cal X}_{k'}^{0,1}(e_j)e^{
    \ii(k\lambda_i+k'\lambda_j - k\varpi_i - k'\varpi_j)
    }~.
    \label{eq:indirect:term2}
\end{multline}
Indirect terms arising from Term 2 will contribute to the amplitudes of
Equation \eqref{eq:disturbing_function:Hint_fourier} for cosine arguments
satisfying 
\begin{equation}
    \pm\pmb{k}\cdot\pmb{\theta}_{i,j} = k'\lambda_j + k\lambda_i + (1-k)\varpi_i +  (1-k')\varpi_j  -2 \Omega_i  + m (\Omega_i-\Omega_j)
\end{equation}
where $m=0,1,2$ and $k,k'\ne0$.
The corresponding indirect amplitudes are given by 
\begin{eqnarray}
    \left[\tilde{C}_{\pmb{k}}^{(\nu_1,\nu_2,\nu_3,\nu_4)}(\alpha)\right]_\mathrm{ind.}
     &=& 
   -\frac{1}{\sqrt{\alpha}}
   {\cal N}^{0,1}_{k',\nu_4}
   {\cal N}^{0,1}_{k,\nu_3}
   (-1)^{\nu_1+\nu_2}
   \binom{\frac{m}{2}}{\nu_1}
   \binom{1 - \frac{m}{2}}{\nu_2}
   (1 +\delta_{m,1} )
\end{eqnarray}
Term 2 contributes indirect terms to the cosine amplitude
$\tilde{C}_{(p,q-p,p-q+1,1-p,m-2,-m)}^{\mathbf{\nu}}$ associated with an
$p:p-q$ resonance for $(k',k)=(p,q-p)$ and the amplitude
$\tilde{C}_{(p,q-p,p-q-1,-1-p,2-m,m)}^{\mathbf{\nu}}$ for $(k',k)=(-p,p-q)$.

\section{Relating orbital-element and canonical-variable expansions}
\label{app:canonical_expansion}

In this appendix we derive expressions for the coefficients ${C}_{\bf
k}^{\pmb{\nu},\pmb{l}}(\alpha)$, expressing the expansion of the disturbing
function in terms of canonical variables, in terms of the coefficients
$\tilde{C}_{\bf k}^{\pmb{\nu}}(\alpha)$ that express the disturbing function's
expansion in orbital elements.  In order to do so, we proceed in two steps:
first, in section \ref{sec:XY_expand}, we express the disturbing function
expansion in terms of the complex variables $X_i,X_j, Y_i$ and $Y_j$, which are
related to \textsf{celmech}'s default set of canonical variables according to
$X_i=\sqrt{\frac{1}{\Lambda_{i,0}}}(\kappa_i-\ii \eta_i)$ and
$Y_i=\frac{1}{2}\sqrt{\frac{1}{\Lambda_{i,0}}}(\sigma_i-\ii \rho_i)$.  Then, in
Section \ref{sec:delta_expand} we expand the resulting expressions from section
\ref{sec:XY_expand} in terms of
$\delta_i=(\Lambda_i-\Lambda_{i,0})/\Lambda_{i,0}$ and $\delta_j$.

\subsection{Relating $z$ and $\zeta$ expansion to $X$ and $Y$ expansion}
\label{sec:XY_expand}

Let us define the new variables $\hat{X}_i\equiv (1+\delta_i)^{1/2}X_i$ and
$\hat{Y}_i\equiv (1+\delta_i)^{1/2}Y_i$ so that $z_i =
\hat{X}_i\paren{1-\frac{1}{4}|\hat{X}_i|^2}^{1/2}$ and $\zeta_i =
\hat{Y}_i\paren{1-\frac{1}{2}|\hat{X}_i|^2}^{-1/2}$ (see Equations
\eqref{eq:z_in_XY} and \eqref{eq:zeta_in_XY}) along with new disturbing
function expansion coefficients $\hat{C}_{\bf k}^{\pmb{\nu}}(\alpha_{ij})$ that
satisfy
\begin{multline}
        \fracbrac{1}{a_j}
        z_i^{k_3}z_j^{k_4}\zeta_i^{k_5}\zeta_j^{k_6}
        \sum_{\pmb{\nu}=0}^\infty 
         \tilde{C}_{\bf k}^{\pmb{\nu}}(\alpha_{ij})
          |\zeta_i|^{2\nu_1}|\zeta_j|^{2\nu_2}|z_i|^{2\nu_3}|z_j|^{2\nu_4}
          =\\
        \fracbrac{1}{a_{j}}
        \hat{X}_i^{k_3}\hat{X}_j^{k_4}\hat{Y}_i^{k_5}\hat{Y}_j^{k_6}
        \sum_{\pmb{\nu}=0}^\infty 
        \hat{C}_{\bf k}^{\pmb{\nu}}(\alpha_{ij})
        |\hat{Y}_i|^{2\nu_1}|\hat{Y}_j|^{2\nu_2}|\hat{X}_i|^{2\nu_3}|\hat{X}_j|^{2\nu_4}
        \label{eq:hamiltonian:poincare:complex_expansion2}~
\end{multline}
where we assume here and throughout this appendix that $k_m\ge0$ for
$m=3,...,6$.\footnote{Whenever any $k_m<0$, the appropriate replacements
$Z^{k_m}\rightarrow \bar{Z}^{-k_m}$ should be made, where $Z$ stands for one of
$X_i,X_j,Y_i$ or $Y_j$.} In order to obtain explicit expressions for the
coefficients $\hat{C}_{\bf k}^{\pmb{\nu}}(\alpha_{ij})$, we write the monomial
$z_i^{k_3}\zeta_i^{k_5}|\zeta_i|^{2\nu_1}|z_i|^{2\nu_3}$ in terms of variables
$\hat{X}_i$ and $\hat{Y}_i$ as 
\begin{eqnarray}
          z_i^{k_3}
          \zeta_i^{k_5}
          s_i^{2\nu_1}
          e_i^{2\nu_3}
          &=&
          \hat{X}_i^{k_3}
          \hat{Y}_i^{k_5}
          |\hat{Y}_i|^{2\nu_1}
          |\hat{X}_i|^{2\nu_3}
          \paren{1-\frac{1}{4}|\hat{X}_i|^2}^{|k_3|/2+\nu_3}
          \paren{1-\frac{1}{2}|\hat{X}_i|^2}^{-|k_5|/2-\nu_1}
          \nonumber\\&\equiv&
          \hat{X}_i^{k_3}
          \hat{Y}_i^{k_5}
          |\hat{Y}_i|^{2\nu_1}
          |\hat{X}_i|^{2\nu_3}
          \sum_{n=0}^\infty {\cal T}^{(n)}_{|k_3|/2+\nu_3,|k_5|/2+\nu_1}|\hat{X}_i|^{2n}
          \label{eq:expansion_double_sum}
\end{eqnarray}
where the coefficients 
\begin{equation*}
{\cal T}^{(n)}_{p,q} = 
    \left(-\frac{1}{2}\right)^{n}
    \sum_{l=0}^n 
    \binom{p}{l}
    \binom{-q}{n-l}
    \paren{\frac{1}{2}}^l
\end{equation*}
are obtained by expanding the factors
$\paren{1-\frac{1}{4}|\hat{X}_i|^2}^{|k_3|/2+\nu_1}$ and
$\paren{1-\frac{1}{2}|\hat{X}_i|^2}^{-|k_5|/2-\nu_3}$ in the first line of
Equation \eqref{eq:expansion_double_sum} and collecting terms $\propto
|\hat{X}_i|^{2n}$. Using Equation \eqref{eq:expansion_double_sum}, we can write
the individual terms appearing in the sum on the left-hand side of Equation
\eqref{eq:hamiltonian:poincare:complex_expansion2} in terms of the variables
$\hat{X}_i,\hat{Y}_i,\hat{X}_j$ and $\hat{Y}_j$ as
\begin{multline}
    \tilde{C}_{\bf k}^{\pmb{\nu}}(\alpha_{ij})
    z_i^{k_3}z_j^{k_4}\zeta_i^{k_5}\zeta_j^{k_6}
    s_i^{2\nu_1}s_j^{2\nu_2}e_i^{2\nu_3}e_j^{2\nu_4} 
    = \\
    \tilde{C}_{\bf k}^{\pmb{\nu}}(\alpha_{ij})
    \hat{X}_i^{k_3}\hat{X}_j^{k_4}\hat{Y}_i^{k_5}\hat{Y}_j^{k_6}
    |\hat{Y}_i|^{2\nu_1}|\hat{Y}_j|^{2\nu_2}
    \sum_{n,m=0}^{\infty}
    {\cal T}^{(n)}_{|k_3|/2+\nu_3,|k_5|/2+\nu_1}
    {\cal T}^{(m)}_{|k_4|/2+\nu_4,|k_6|/2+\nu_2}
    |\hat{X}_i|^{2n}|\hat{X}_j|^{2m}~.
    \label{eq:zzeta_ij_in_XY_ij}
\end{multline}
Collecting terms in Equation \eqref{eq:zzeta_ij_in_XY_ij} with $N_3 = \nu_3 +
n$ and $N_4 = \nu_4 + n$, we arrive at the following expression relating
$\hat{C}_{\bf k}^{\pmb{\nu}}(\alpha_{ij})$ to the coefficients $\tilde{C}_{\bf
k}^{\pmb{\nu}}(\alpha_{ij})$:
\begin{equation}
	\hat{C}_{\bf k}^{(N_1,N_2,N_3,N_4)}(\alpha_{ij})=
	\sum_{n=0}^{N_3}
	\sum_{m=0}^{N_4}
	{\cal T}^{(n)}_{|k_3|/2+\nu_3,|k_5|/2+\nu_1}
	{\cal T}^{(m)}_{|k_4|/2+\nu_4,|k_6|/2+\nu_2}
	\tilde{C}_{\bf k}^{(N_1,N_2,n,m)}(\alpha_{ij}).
	\label{eq:Chat_in_term_of_C}
\end{equation}

\subsection{Expansion in $\delta_i$ and $\delta_j$}
\label{sec:delta_expand}
To derive explicit expressions for the coefficients ${C}_{\bf
k}^{\pmb{\nu},(l_1,l_2)}$ in terms of $\hat{C}_{\bf k}^{\pmb{\nu}}$, we write
out the $\delta_i$ and $\delta_j$ dependence of the terms appearing in the sum
on right-hand side of Equation \eqref{eq:Chat_in_term_of_C} explicitly as 
\begin{eqnarray}
    \fracbrac{1}{a_{j}}
        \hat{C}_{\bf k}^{\pmb{\nu}}(\alpha_{ij})
        &&
        \hat{X}_i^{k_3}\hat{X}_j^{k_4}\hat{Y}_i^{k_5}\hat{Y}_j^{k_6}
        |\hat{Y}_i|^{2\nu_1}|\hat{Y}_j|^{2\nu_2}|\hat{X}_i|^{2\nu_3}|\hat{X}_j|^{2\nu_4}
        \nonumber\\&=&
    \left\{\frac{1}{a_{j,0}}
    \paren{1+\delta_i}^{-(|k_3|+|k_5|)/2-\nu_1 - \nu_3}
    \paren{1+\delta_j}^{2-(|k_4|+|k_6|)/2-\nu_2 - \nu_4}
    \hat{C}_{\bf k}^{\pmb{\nu}}
    \paren{
        \alpha_{ij,0}
        \fracbrac{1+\delta_i}{1+\delta_j}^2
    }\right\}
    \nonumber\\
    &&\times
    {X}_i^{k_3}{X}_j^{k_4}{Y}_i^{k_5}{Y}_j^{k_6}
        |{Y}_i|^{2\nu_1}|{Y}_j|^{2\nu_2}|{X}_i|^{2\nu_3}|{X}_j|^{2\nu_4}~.
\end{eqnarray}
Accordingly, we may write
\begin{eqnarray}
    C_{{\bf k}}^{\pmb {\nu},(l_1,l_2)}(\alpha_{ij,0}) 
    = \frac{1}{l_1!l_2!}
    \pd{
        ^{l_1+l_2}
    }{
        \delta_i^{l_1}\partial\delta_j^{l_2}
    }
    \left[
        (1+\delta_i)^{-p_i/2}
        (1+\delta_j)^{-p_j/2}
        \hat{C}_{{\bf k}}^{\pmb{\nu}}\left(
            \alpha_0
            \fracbrac{
                1+\delta_i
                }{
                1+\delta_j
            }^2
            \right)
    \right]_{(\delta_i,\delta_j)=0}
    \label{eq:Cll_in_terms_of_Chat}
\end{eqnarray}
where $p_i=|k_3| + |k_5| + 2\nu_1  +2\nu_3$ and $p_j=4  + |k_4| + |k_6| +
2\nu_2  +2\nu_4$. For the case $l_1=l_2=0$, from Equation
\eqref{eq:Cll_in_terms_of_Chat} we simply have $C_{{\bf k}}^{\pmb
{\nu},(0,0)}(\alpha_{ij,0}) = \hat{C}_{{\bf k}}^{\pmb{\nu}}(\alpha_{ij,0})$.
More generally, an explicit expressions of $C_{{\bf k}}^{\pmb
{\nu},(l_1,l_2)}(\alpha_{ij,0})$ in terms of $\hat{C}_{{\bf k}}^{\pmb{\nu}}$
and its derivatives can be derived by applying the product rule for derivatives
along with di Bruno's formula \citep[e.g.,][]{johnson2002curious} generalizing
the chain rule to higher derivatives to Equation
\eqref{eq:Cll_in_terms_of_Chat}.  A tedious but straightforward calculation
yields
\begin{eqnarray}
     C_{\bf{k}}^{\pmb{\nu},(l_1,l_2)}(\alpha_0)
    &=&
    \sum_{m_1=0}^{l_1}
    \sum_{r_1=1}^{l_1-m_1}
    \sum_{m_2=0}^{l_2}
    \sum_{r_2=1}^{l_2-m_2}
    \Psi_{l_1,l_2,p_1,p_2,m_1,m_2,r_1,r_2}
    \alpha_0^{(r_i+r_j)}
    \dd{^{(r_1+r_2)}}{\alpha^{(r_1+r_2)}}
    \hat{C}_{\bf{k}}^{\pmb \nu}(\alpha)\bigg|_{\alpha=\alpha_0}\nonumber
\end{eqnarray}
where 
\begin{equation*}
    \Psi_{l_1,l_2,p_1,p_2,m_1,m_2,r_1,r_2}
    =
    \frac{1}{l_1!l_2!}
    \binom{l_2}{m_2}
    \binom{l_1}{m_1}
    (-2r_1-p_2/2)_{m_2}
    (-p_1/2)_{m_1}
    {\cal B}_{l_1-m_1,r_1}^{(+)}
    {\cal B}_{l_2-m_2,r_2}^{(-)}~,
\end{equation*}
    $(x)_k \equiv x(x-1)...(x-k+1)$,
    and
\begin{equation*}
    {\cal B}_{k,n}^{(\pm)}= B_{n,k}\left(\binom{\pm2}{1}1!,...,\binom{\pm2}{n-k+2}(n-k+2)!\right)
\end{equation*} with $B_{n,k}$ denoting partial Bell polynomials.
\bibliography{main}{}
\bibliographystyle{aasjournal}
\end{document}